\documentclass{aa}  

\usepackage{graphicx}
\usepackage{txfonts}
\usepackage{tabularx} 
\usepackage{longtable}
\usepackage{array}
\usepackage[switch]{lineno}
\newcolumntype{C}[1]{>{\centering\arraybackslash}m{#1}}
\usepackage{lscape}
\usepackage[flushleft]{threeparttable} 
\usepackage{caption}
\def\degree{${}^{\circ \  }  $}
\def\twco{$^{12}$CO }
\def\thco{$^{13}$CO }

\usepackage{hyperref}

\begin{document}

   \title{Large-field CO (J=1-0) observations toward SNR G150.3+4.5}


   \author{Jian-Cheng Feng, \inst{1,2}
           Xuepeng Chen, \inst{1,2}
           Yang Su, \inst{1,2}
           Li Sun, \inst{1,2}
           Shiyu Zhang, \inst{1,2}
           Xin Zhou, \inst{1,2}
           \and 
           Weihua Guo \inst{3,1}
          }

   \institute{Purple Mountain Observatory, Chinese Academy of Sciences, No.10 Yuanhua Road, Qixia District, Nanjing 210023, China\\
              \email{fengjc@pmo.ac.cn, xpchen@pmo.ac.cn}
         \and
             University of Science and Technology of China, No.96, JinZhai Road Baohe District,Hefei,Anhui, 230026, China
        \and 
            Institute for Theoretical Physics and Cosmology, Zhejiang University of Technology, Hangzhou 310023, China\\
             }

   \date{Received 2024 March 26; accepted 2024 March 28.}

 
  \abstract
{}
{
We aim to investigate the molecular environment of the supernova remnant (SNR) G150.3+4.5, and explore its association with ambient molecular clouds (MCs).
}
{
We present large-field CO (J=1-0) molecular line observations toward SNR G150.3+4.5, using the 13.7 m millimeter telescope of the Purple Mountain Observatory. The observations have an angular resolution of $\sim 55 ''$. 
We analyzed the spatial distribution of MCs in relation to the SNR shell detected in previous Urumqi $\lambda$ 6 cm radio observations and examined the CO spectra for kinematics information.
}
{
We find that MCs at the velocity range of [-14, -2] km s$^{-1}$ are spatially distributed along the radio shell of the SNR. Line broadening and asymmetries are observed in the CO spectra of the clouds. Moreover, we find that the molecular clouds around the shell have systematic velocity gradients in the position-velocity (PV) diagrams. Both morphology alignment and gas kinematics suggest that the SNR is associated with the ambient MCs at $\sim$ 740 pc. Based on the CO gas distance, the dimension and the age of the SNR is estimated to be 40 pc $\times$ 33 pc and 3.8 $ \times 10^4$ years, respectively. The very high energy emission of 1LHAASO J0428+5531 toward the SNR may originate from the interaction between the SNR and the surrounding MCs.
}
{
}


   \titlerunning{Large-field CO (J=1-0) observations toward SNR G150.3+4.5}
    \authorrunning{Jian-Cheng Feng et al.}
       \maketitle
%

\section{Introduction}

The explosion of a supernova (SN) has a substantial impact on the physical and chemical properties of interstellar medium (ISM) in the Galaxy. Core-collapse supernovae (SNe) are thought to be commonly located in the vicinity of molecular clouds (MCs), due to the short lifetime of their progenitors (i.e., massive stars) after being formed in giant MCs (see \citealt{huang1986molecular}). Therefore, the subsequent supernova remnants (SNRs) may interact with the natal MCs during their evolution (see, e.g., a review by \citealp{dubner2015radio}).
SNRs significantly impact the lifecycle of molecular clouds and the efficiency of star formation by injecting substantial momentum into them \citep{kim2015momentum,chevance2023life}: Negative feedback, where SNRs displace molecular material and suppress star formation \citep{kortgen2016supernova, kruijssen2019fast,sano2023alma}, contrasts with positive feedback, wherein SNRs compress gas to trigger new star formation \citep{inutsuka2015formation,miret2022star}.

For now, there are more than 80 Galactic SNRs confirmed or suggested to be in physical association with MCs among $\sim$\,300 known SNRs in the Milky Way (see, e.g., \citealp{jiang2010cavity,kilpatrick2015systematic,zhou2020molecular}).
\cite{jiang2010cavity} summarized six kinds of multi-wavelength observational evidence for judging the association/interaction between SNR and MCs. Firstly, if MCs overlapping with an SNR show morphological agreement or correspondence between molecular features and SNR features (e.g., radio shells), it can be considered as a candidate of the SNR-MC association system. Yet, it is not strong evidence because of the overlap of multiple MCs in different velocity ranges along the line of sight. More observational evidence, such as OH 1720 MHz maser emission (e.g., \citealt{goss1968oh,reynoso2000co,dubner2004molecular}, etc.), molecular line broadening (e.g., \citealt{denoyer1979discovery,kilpatrick2015systematic,zhou2023systematic}, etc.), line emission with a high high-to-low excitation line ratio (e.g., \citealt{seta1998enhanced}, etc.), and other shock-gas interaction signatures (e.g., \citealt{wootten1981dense}, etc.), etc, are needed to further verify the association between SNRs and MCs. In a recent study by \cite{zhou2023systematic}, a comprehensive search for kinematic and spatial correlations between SNRs and MCs in the Milky Way was conducted. The study suggested that as many as 80\% of SNRs could be associated with MCs.

Large-field observations of molecular lines provide physical, chemical, and dynamic information about molecular gas, which is essential to study the interaction between SNRs and MCs, particularly for those remnants with large angular sizes. This kind of observations can not only help us to better understand the nature of the SNRs, but also to improve our knowledge about stellar feedback (e.g., stellar wind and SNe) on the MCs, as well as the cycling of the ISM in the galaxy. Furthermore, SNRs have been suggested as potential sources of Galactic cosmic rays (CRs) for a long time (e.g., \citealp{baade1934cosmic}). Recent high-energy observations indicate that middle-aged SNRs interacting with molecular clouds can emit hadronic $\gamma$-rays. For example, resolved images from VERITAS \citep{acciari2009observation} and Fermi \citep{abdo2010observation} presented early insights into the morphological correspondence of high-energy $\gamma$-ray emissions with the dense shocked molecular clump in IC443, which provide us the ability to pinpoint TeV emission to individual sites and help us constrain the origin of CRs with the environment where they originate. Subsequent studies employed various methods to examine the interaction of shock waves with molecular gas, including the detection of H$_2$ (e.g., \citealt{reach2019supernova}), the analysis of large velocity gradients (LVG) (e.g., \citealt{dell2020interstellar}), and the observations of molecular line shock tracers (e.g., \citealt{cosentino2019interstellar,cosentino2022negative}). These multi-wavelength studies have significantly advanced our comprehension of the mechanisms behind CR acceleration and propagation in the complex interplay between SNRs and the surrounding molecular gas.

The Milky Way Imaging Scroll Painting (MWISP) project\footnote{\href{http://www.radioast.nsdc.cn/mwisp.php}{http://www.radioast.nsdc.cn/mwisp.php}} is a large unbiased CO\,(J = 1-0) multi-line survey toward the northern Galactic plane, using the 13.7\,m millimeter telescope of the Purple Mountain Observatory (PMO; \citealp{su2019milky,sun2021examinations}). The systematic survey provides high-quality of large filed of view data for studying the SNR-MC interaction in the Milky Way (see, e.g., \citealp{su2017hess,su2018large,chen2017large,zhou2020molecular,zhou2023systematic}).

As part of the MWISP survey, we present in this work large-field CO\,(J = 1-0) observations toward the recently discovered SNR G150.3+4.5 \citep{gerbrandt2014search, gao2014discovery}. The SNR has a loop structure with three major shells. It was first reported as a candidate SNR in the Canadian Galactic Plane Survey (CGPS) by \cite{gerbrandt2014search}, in which only the southeastern shell was observed (called G150.8+3.8). A thin filament was also found in the DSS optical map, which spatially coincides with the radio shell \citep{gerbrandt2014search}. Taking advantage of the large-field Urumqi $\lambda$ 6\,cm survey, \cite{gao2014discovery} reported the SNR as a roughly complete radio loop, with a size of 3\fdg0\,$\times$\,2\fdg5. The southeastern shell is the most prominent one, curving to the centre of the lower south. The fainter southwestern shell is curving to the centre of the lower south as well. 

Additionally, an extended source from the Second Catalog of Hard Fermi-LAT Sources (2FHL) J0431.2+5553e was identified in proximity to this SNR, with a radius of r = 1\fdg27\,$\pm$\,0\fdg04 \citep{cohen2016gamma}. This source is also listed in the Fourth Fermi-LAT catalog as 4FGL J0427.2+5533e \citep{thompson2019fourth}. Using more than 10\,yr of Fermi-LAT data, \cite{devin2020high} investigated the morphological and spectral properties of the $\gamma$-ray emission towards the SNR from 300\,MeV to 3\,TeV. More recently, an even high-energy ($>$\,25\,TeV) extended source was detected toward the remnant by the Kilometer Square Array (KM2A) and Water Cherenkov Detector Array (WCDA) at the Large High Altitude Air Shower Observatory (LHAASO; \citealp{cao2023lhaaso}).

Based on the MWISP CO data, we investigate the interstellar MCs toward the SNR. We describe the observations and data reduction in Sect. 2. Observational results are presented in Sect. 3 and discussed in Sect. 4. The main conclusions of this study are summarized in Sect. 5.


\section{OBSERVATIONS AND DATA REDUCTION} \label{sec:obs}

The observations of CO (J = 1-0) toward SNR G150.3+4.5 were performed from 2012 to 2022 with the PMO 13.7m telescope at Delingha in China. The nine-beam Superconducting Spectroscopic Array Receiver (SSAR;  \citealt{shan2012development}) was working as the front end in sideband separation mode. Three CO (J = 1-0) lines were simultaneously observed, $^{12}$CO at the upper sideband (USB), while two other lines $^{13}$CO and C$^{18}$O at the lower sideband (LSB). 
Typical system temperatures were around 210 K for the USB and around 130 K for LSB, and the variations among different beams are less than 15\%. 
A Fast Fourier Transform (FFT) spectrometer with a total bandwidth of 1 GHz and 16384 channels was used as the back end.  The corresponding velocity resolutions were 0.16 $\rm km \ s^{-1}$ for the $^{12}$CO line and 0.17 $\rm km \ s^{-1}$ for both $^{13}$CO and C$^{18}$O. 
The observation area was segmented into individual 30$'$ $\times$ 30$'$ cells, each mapped using the on-the-fly (OTF) mode. Each cell was mapped at least twice, along the Galactic longitude and latitude, to reduce scanning effects. The half-power beamwidth (HPBW) was $\thicksim$ 52$''$ for the $^{12}$CO line and $\thicksim$ 55$''$ for both the $^{13}$CO and C$^{18}$O lines. The pointing accuracy was $\thicksim$ 5$''$. 
Antenna temperatures ($T_{\rm A}$) were calibrated using the standard chopper-wheel method, and the main-beam temperatures ($T_{\rm mb}$) were derived using the formula $T_{\rm mb}=T_{\rm A} /B_{\rm eff}$, where the main-beam efficiencies ($B_{\rm eff}$) were approximately 44\% for USB and 48\% for LSB. Calibration errors were estimated to be within 10\%. 

After the removal of bad channels and abnormal spectra, along with corrections for first-order (linear) baseline fitting, the data were regridded into standard FITS files with a pixel size of 30$''$ × 30$''$, approximately half of the beam size. The mean rms noise level of all final spectra was about 0.5 K for $^{12}$CO and $0.3 \ \rm K$ for $^{13}$CO and C$^{18}$O. Velocities were referenced to the local standard of rest (LSR). All the CO data were reduced by using the GILDAS/CLASS package\footnote{\href{https://www.iram.fr/IRAMFR/GILDAS/}{https://www.iram.fr/IRAMFR/GILDAS/}}.


\section{Results} 
\label{sec:results}

\subsection{Overview of the Molecular Gas Toward SNR G150.3+4.5}  
\label{subsec:overview}

\subsubsection{Morphology and Spectra}

\begin{figure*}
    \centering
    \includegraphics[width=1\linewidth,clip]{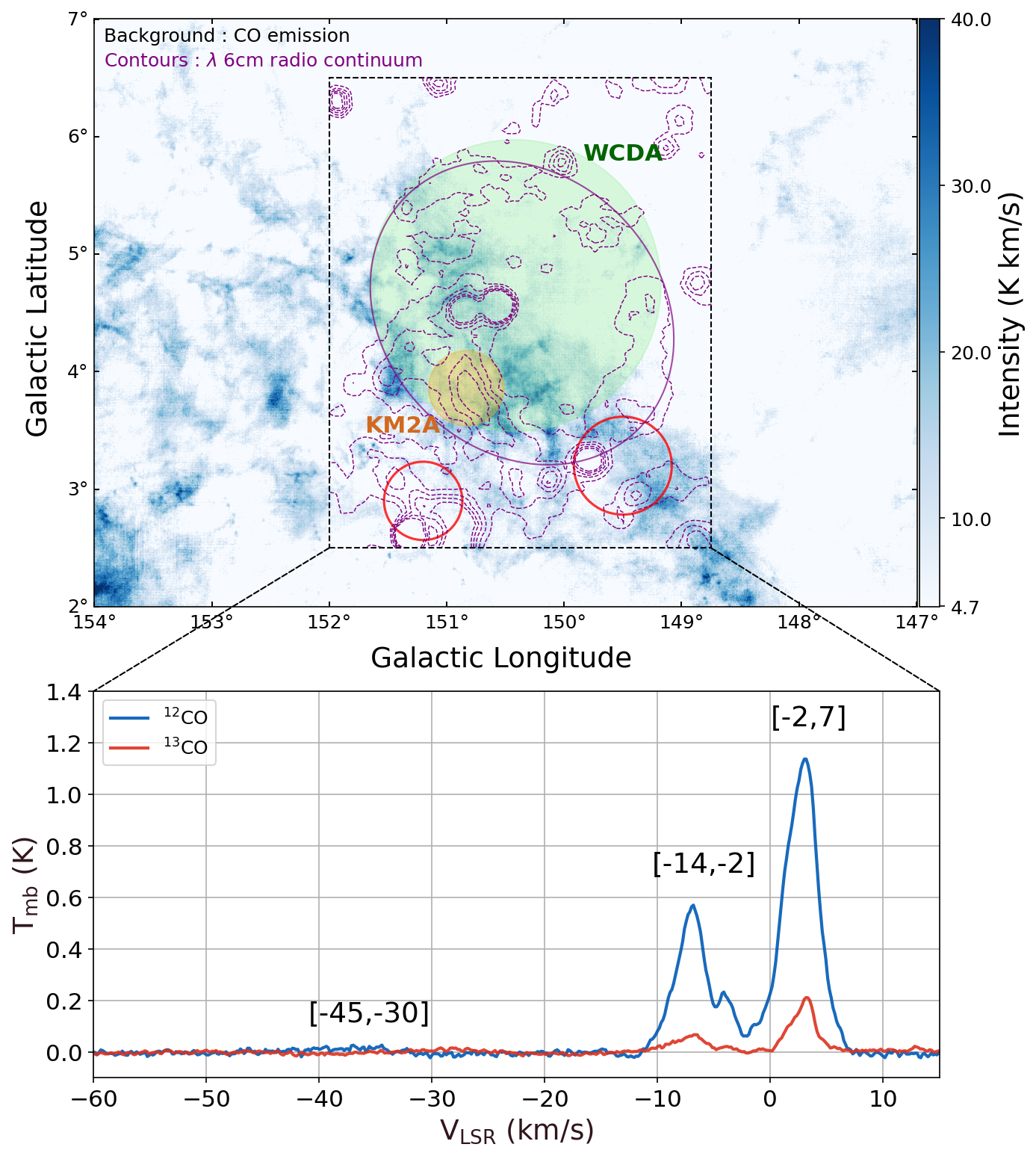}
    \caption{A large-field global view toward SNR G150.3+4.5. The upper panel shows an integrated intensity map derived from the \twco emission with a velocity range between -60 and 15 $\rm km \ s^{-1}$ ($3\sigma = 4.7 \, \text{K km s}^{-1}$). The purple dashed contours represent the Urumqi $\lambda$  6 cm radio continuum emission \citep{gao2014discovery}, and the purple ellipse shows the size of the remnant in the 6 cm radio observations. The contours run at $3.0 + n \times 3.0\,\text{mK}\,(n = 1, 2, \ldots 5)$. The orange and green circles represent the very high-energy source 1LHAASO J0428+5531 \citep{cao2023lhaaso}, with orange indicating KM2A component and green denoting WCDA component. The red circles represent two SNR candidates G151.2+2.9 \citep{kerton2007sharper} and G149.5+3.2 \citep{gerbrandt2014search}. The bottom panel depicts the average spectra of the regions marked in the upper panel with the rectangle, with the spectra of $\rm ^{12}CO$ (blue) and $\rm ^{13}CO$ (red). The annotations on the spectra pinpoint the velocity ranges of individual components.}
    \label{fig:global view}
\end{figure*}

\begin{figure*}
\centering
\includegraphics[width=16.4cm,clip]{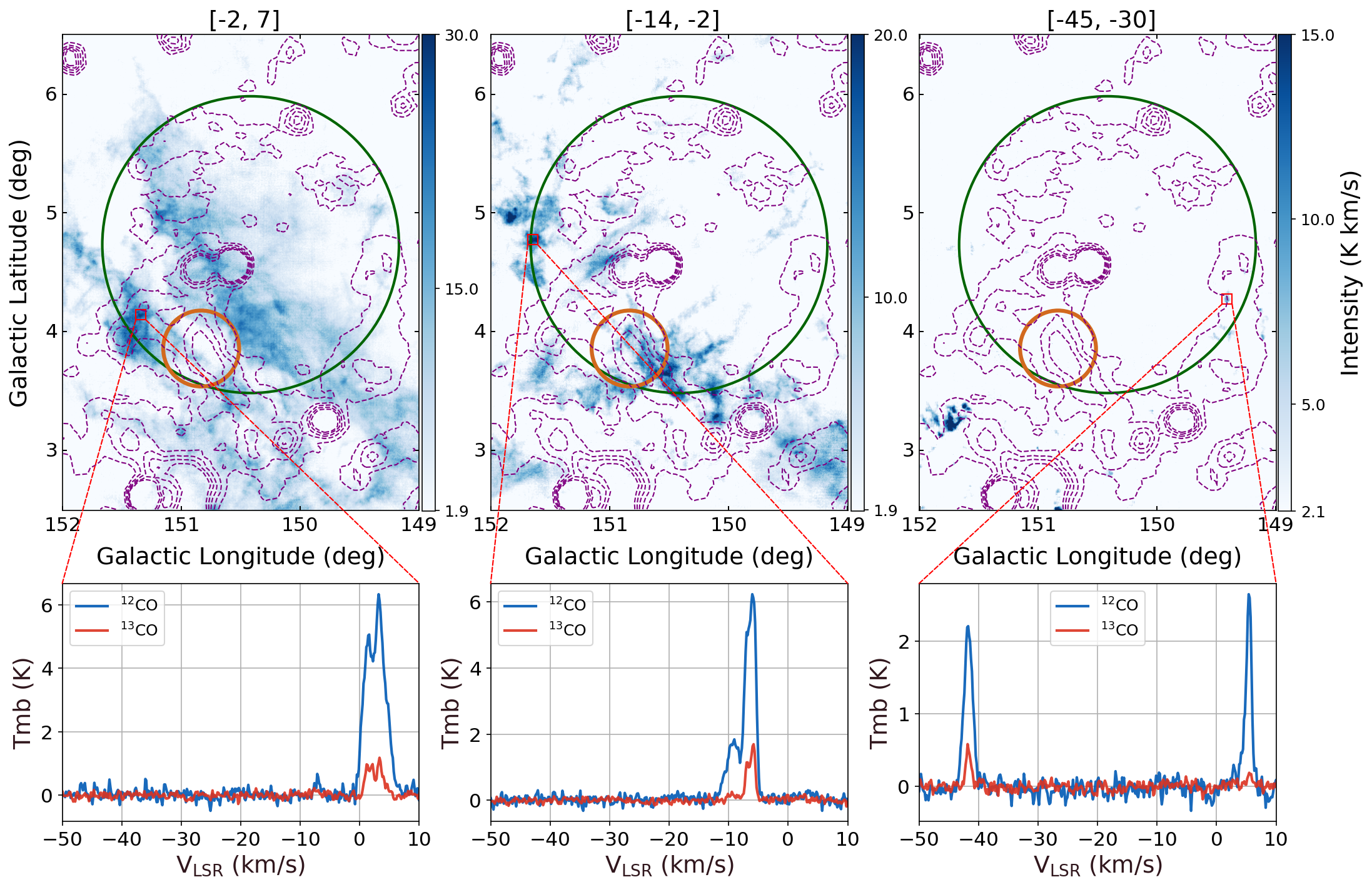}
\caption{The velocity-integrated intensity image with typical spectra over three individual velocity ranges, left: [-2, 7], mid: [-14, -2], and right: [-45, -30] $\rm km \ s^{-1}$. The top panel of each image displays intensity maps derived from CO observations. The color scale ranges from $3\sigma$, as indicated on the color bar. The purple dashed contours represent Urumqi $\lambda$  6 cm radio continuum emission same as Fig.\ref{fig:global view}. The orange and green circles represent the LHAASO source 1LHAASO J0428+5531, with orange indicating KM2A components and green denoting WCDA components. The lower panel shows the average spectra for the highlighted region, where the blue and red spectra represent $\rm ^{12}CO$ and $\rm ^{13}CO$, respectively.}
\label{Fig:allvelocity}
\end{figure*}

In Fig. \ref{fig:global view}, we present the large-field CO emission towards the entire region of SNR G150.3+4.5. The blue color map shows the integrated intensity of \twco with a velocity range from $-60$ to 15 $\rm km \ s^{-1}$. The black dashed rectangle region illustrates the location of the SNR, which covers the entire radio loop of the SNR observed in the Urumqi $\lambda$ 6 cm survey \citep{gao2014discovery}. 
We overlay contours and circles from observations at other wavelengths to examine the morphological alignment between CO emissions and the the SNR.
In the Urumqi $\lambda$ 6 cm radio continuum observations, the SNR appears as a complete loop structure as the purple dashed contours shown in Fig. \ref{fig:global view}. In the high-energy observations at the LHAASO \citep{cao2023lhaaso}, the WCDA (1 TeV to 25 TeV) component aligns closely with this radio loop, while the KM2A ($>$ 25 TeV) component of 1LHAASO J0428+5531 is situated in the southwest of the SNR. The CO emission is distributed throughout the SNR. To study the relationship between molecular gas and the SNR, we first try to distinguish different components of the molecular gas in this region along the line of sight in the velocity space.

We present the average spectra of \twco and \thco within the rectangular region in the lower panel of Fig. \ref{fig:global view}. We find that there are mainly three CO velocity components: [-2, 7] $\rm km \ s^{-1}$, [-14, -2] $\rm km \ s^{-1}$, and [-45, -30] $\rm km \ s^{-1}$. Notably, the CO emission from the [-45, -30] $\rm km \ s^{-1}$ component is relatively faint and is scarcely discernible in the spectra.  To further study the morphology and spatial distribution of these components, we present the velocity-integrated intensity maps of each component in Fig. \ref{Fig:allvelocity}. The blue background in the top panels shows the \twco intensity integrated over the velocity range of each component. The bottom panels display spectra samples of the CO isotopologues from the brightest position within the SNR surrounding in each velocity component. These samples are averaged over areas marked by the red square, measuring 2.5$'$ $\times$ 2.5$'$ (5 pixels $\times$ 5 pixels) on the intensity maps.

We note that the [-2, 7] $\rm km \ s^{-1}$ component corresponds to the giant molecular filament reported by \citet{xiong2017co}. Within this velocity range, the CO emission shows two elongated and parallel structures. \cite{zhou2023systematic} noted that the radio shell overlaps with the molecular gas in this velocity range. 
However, no broad line is identified in this velocity range. The spectra of [-2, 7] $\rm km \ s^{-1}$ component exhibit two peaks for both \twco and \thco. The Gaussian-like shape of the spectra suggests a multi-component molecular complex, rather than the shocked gas in the SNR-MCs interaction event. 
Therefor, we suggest that the [-2, 7] $\rm km \ s^{-1}$ component of the molecular gas is unlikely to be associated with the SNR.

The  [-14, -2] $\rm km \ s^{-1}$ component shows multiple clouds in the intensity map. The southern cloud, notably one of the densest within this component, aligns closely with the distinct southern shell of the Urumqi $\lambda$  6 cm emission. We also detect faint CO emissions at the edges of both the western and northern radio shells. The CO emission, spatially distributed around the radio loop, suggests a morphology that could be relevant to the SNR. Furthermore, we observe non-Gaussian profiles (line broadening or wing features) in the \twco spectra of this component (see Sect. \ref{sec:shocked}), suggesting potential shock disturbances.

As for the [-45, -30] $\rm km \ s^{-1}$ component, we only find some faint CO emission in the western radio shell. The Gaussian-like spectra of this component imply no relation to the SNR.
Based on the CO morphology and spectra analysis of the three velocity components, we suggest that only the [-14, -2] $\rm km \ s^{-1}$ molecular clouds are potentially associated with the SNR.

It should be noted that another two SNR candidates, G151.2+2.9 \citep{gao2014discovery} and G149.5+3.2 \citep{gerbrandt2014search} are located within our observed region and partly overlap with the SNR as shown in Fig. \ref{fig:global view}.
However, we do not identify line broadening features in the spectra within these regions. In this case, we suggest these two SNR candidates are not associated with our identified molecular clouds.


\subsubsection{Distance}
\label{sec:distance}

\begin{figure*}
  \resizebox{\hsize}{!}{\includegraphics{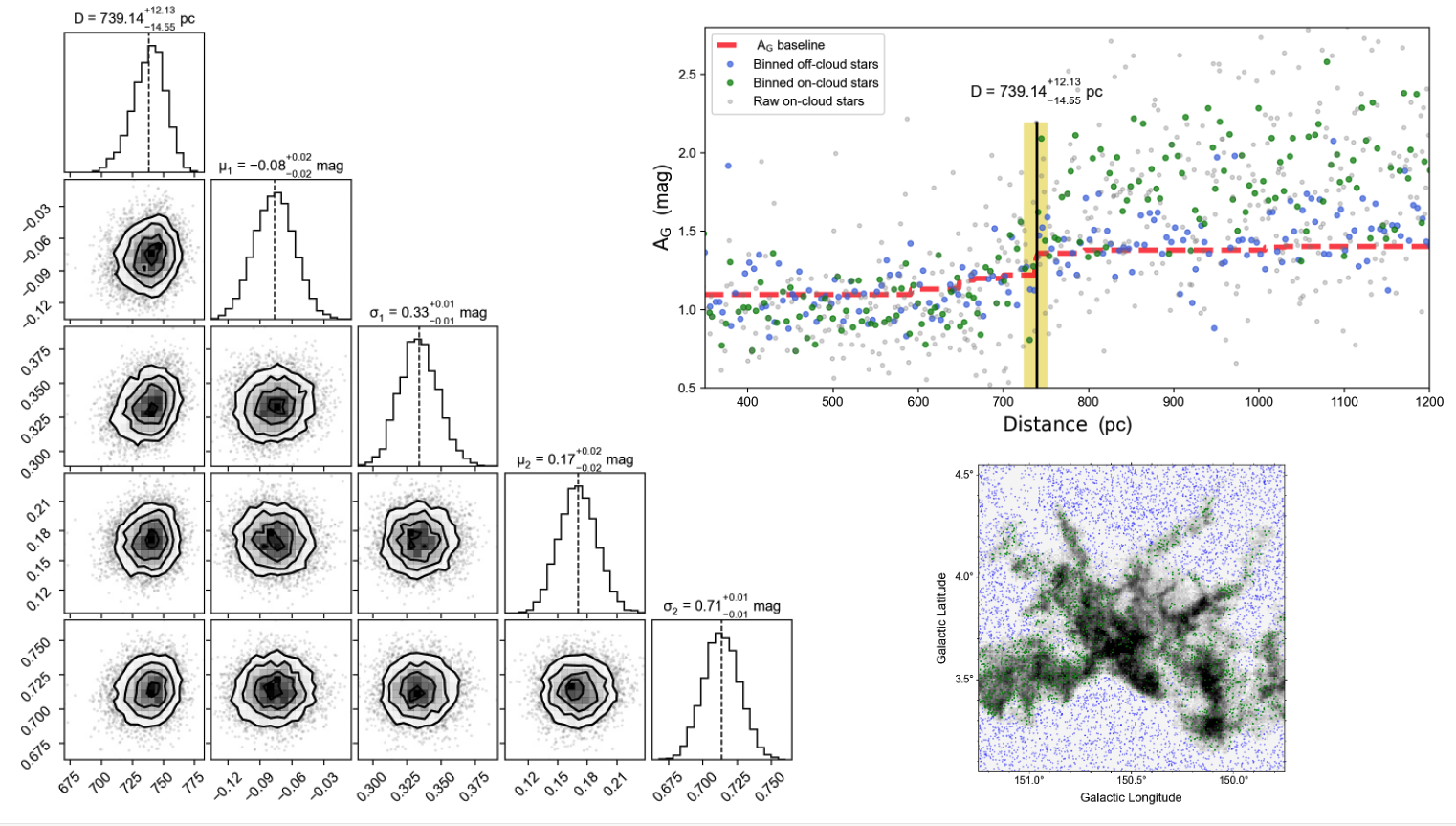}}
\caption{
The distance of the southern molecular clouds. The bottom-right shows the CO velocity-integrated intensity images.
In the top-right panels, the green and blue points present on- and off-cloud stars (binned every 5 pc), respectively. The dashed red lines show the modeled extinction AG. The distances were derived with raw on-cloud Gaia DR3 stars, which are represented with gray points. 
The black vertical lines indicate the distance (D) estimated with Bayesian analyses and Markov Chain Monte Carlo (MCMC) sampling, and the shadow areas depict the 95\% Highest Posterior Density (HPD) range of distances. The corner plots of the MCMC samples are displayed on the left. The mean of the samples is shown with solid lines, and the systematic uncertainty is not included. The distance of the southern molecular cloud is measured to be $739^{+12}_{-14}$ pc.
}
\label{Fig:distance}
\end{figure*}

\begin{figure*}
  \resizebox{\hsize}{!}{\includegraphics{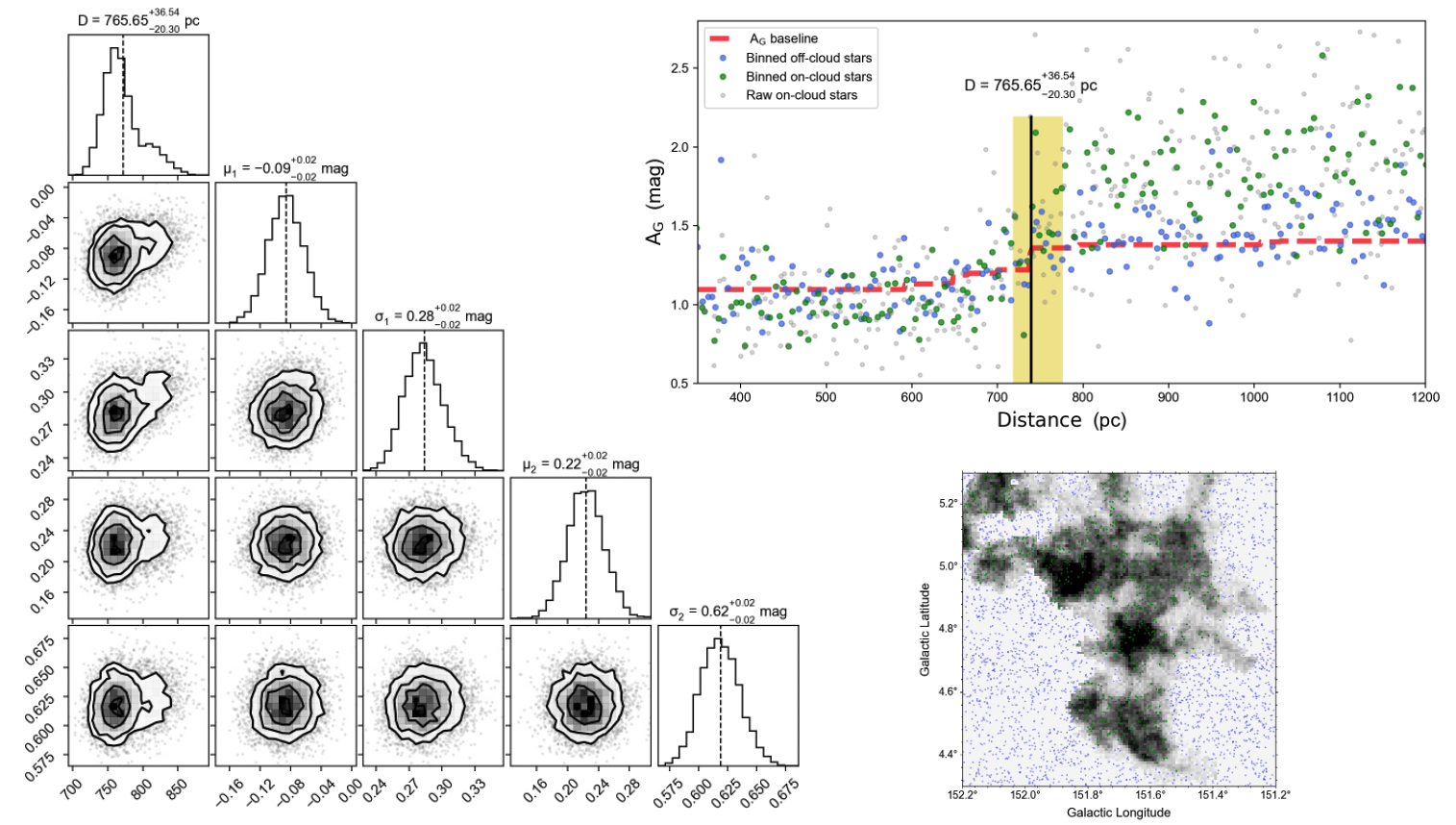}}
\caption{
Same as in Fig. \ref{Fig:distance} but for the eastern cloud. The distance of the eastern molecular cloud is measured to be $765^{+36}_{-20}$  pc. 
}
\label{Fig:distance1}
\end{figure*}


Distance is one of the most important parameters to derive the properties of the gas. Based on the BEEP-II method described in \citet{yan2021distances}, we try to measure the heliocentric distance of molecular clouds from parallax and G band extinction (A$_G$) measurements in the Gaia DR3 \citep{brown2021gaia}. The BEEP-II method is based on the principle that molecular clouds typically exhibit higher optical extinction compared to other phases of the ISM. Utilizing Bayesian analyses, the method derives distances by pinpointing the breakpoint in stellar extinction in the direction of the molecular cloud, termed the “on-cloud” region. To corroborate this breakpoint, the extinction of Gaia stars surrounding the molecular cloud, referred to as the “off-cloud” regions, is also taken into account. The method boasts a systematic error of approximately 5\%, making it a reliable tool for distance estimation.

For the distance of the [-14, -2] $\rm km \ s^{-1}$ component, we first select the southern cloud as the sample for the distance measurement, since it is the largest and also the strongest cloud in this velocity range. The result is shown in Fig. \ref{Fig:distance} and the estimated distance is $739^{+12}_{-14}$  pc. Moreover, we obtain a similar distance at $765^{+36}_{-20}$ pc for another major cloud located in the east of the SNR (see Fig. \ref{Fig:distance1}). Considering the uncertainty, we suggest the two major clouds in this component are at the same distance. We therefore adopt a distance of 740 pc for the [-14, -2] $\rm km \ s^{-1}$ component for further analysis. 

We also apply the same method to estimate the distance for the [-2, 7] $\rm km \ s^{-1}$ component and obtain a distance of approximately 200 pc. 
However, we cannot estimate the distance of the [-45, -30] $\rm km \ s^{-1}$ component using this method. The emission is too faint and the component lies beyond the limit of the BEEP-II method. 
Therefore, we employ the spatial-kinematic method based on the galactic parameters of model A5 in \cite{reid2014trigonometric} to estimate the distance of the [-45, -30] $\rm km \ s^{-1}$ component. We obtain an approximate distance of 3.3 kpc.


\subsection{Properties of the [-14, -2] Molecular Clouds}
\label{sec:properties}

\begin{figure*}[h!]
\centering
\includegraphics[width=1\linewidth,clip]{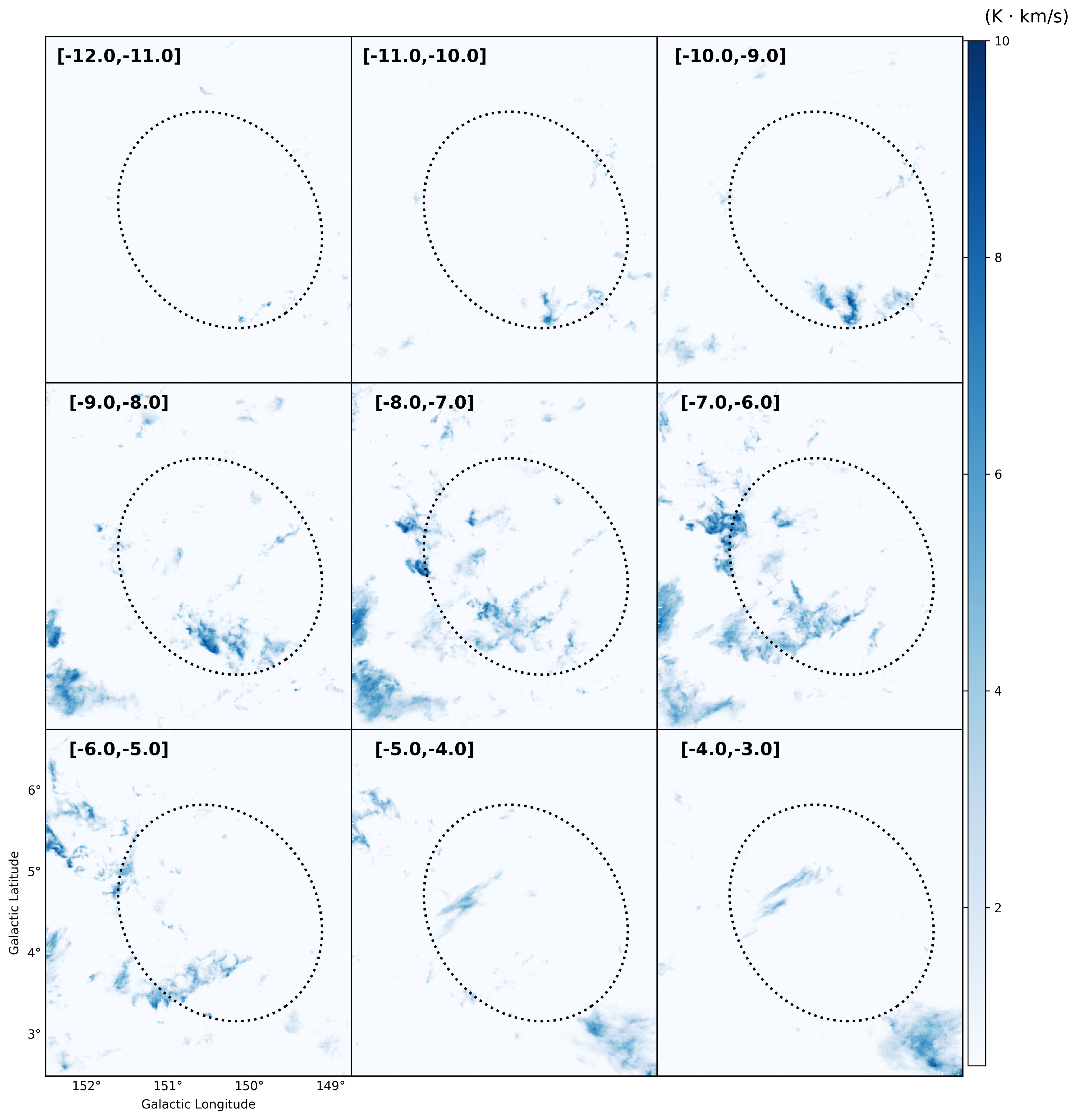}
\caption{ The \twco emission channel maps integrated over each 1 $\rm km\  s^{-1}$  ($3\sigma = 0.54 \, \text{K km s}^{-1}$). Velocities range (in $\rm km\  s^{-1}$) are indicated in the top left corner of each panel. The dotted ellipse shows the size of the remnant in the Urumqi $\lambda$  6 cm radio observations.}
\label{Fig:channel}
\end{figure*}

\begin{figure}[h!]
\includegraphics[width=1\linewidth,clip]{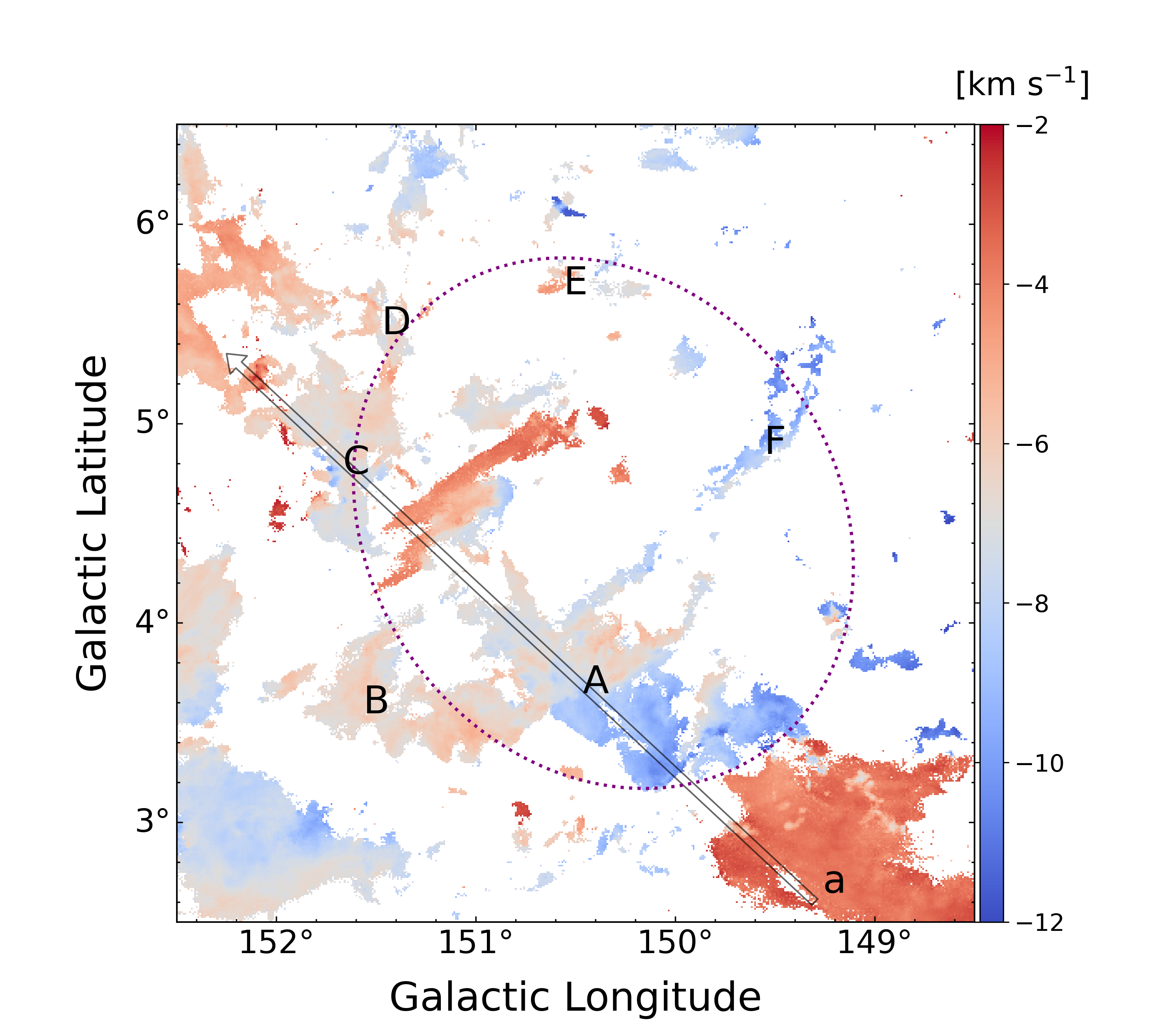}
\caption{ The \twco intensity-weighted velocity map integrated between -14 and -2 $\rm km \ s^{-1}$. The capital letters illustrate the position of the spectra in Fig. \ref{Fig:spec}. The black arrows represent the direction of the position-velocity diagram of slice a in Fig. \ref{Fig:pv}. The purple dashed ellipse shows the size of Urumqi $\lambda$  6 cm radio continuum emission.}
\label{Fig:m1}
\end{figure}

In this section, we focus on the [-14, -2] velocity component, because of its potential for an interaction between the SNR and molecular clouds. Figure \ref{Fig:channel} presents the velocity channel map of the component. The southwestern gas is evident around -10 $\rm km \ s^{-1}$, whereas the eastern gas is prominent at -7 $\rm km \ s^{-1}$, indicating a distinct velocity difference between the east and west. A similar trend is observed in the intensity-weighted velocity map of \twco (shown in Fig. \ref{Fig:m1}). Adopting the heliocentric distance of 740 pc, we can derive the physical properties of the gas structures in the [-14, -2] $\rm km \ s^{-1}$ component.

\subsubsection{Clouds Identification}
\label{sec:clouds}

\begin{figure}[h!]
\resizebox{\hsize}{!}{\includegraphics{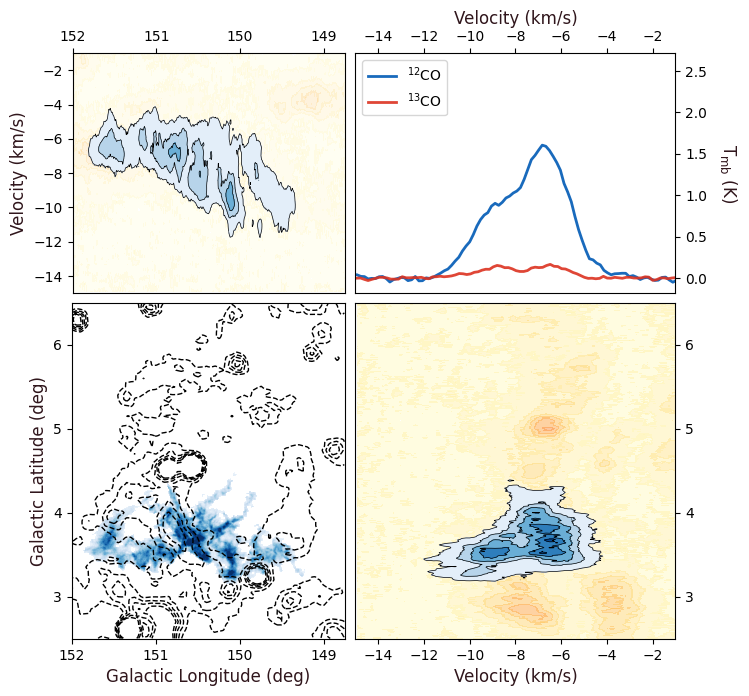}}
\caption{Lower-left: The blue-filled contour is the velocity-integrated intensity image of MC G150.6+03.7 masked by \textsc{DBSCAN} with the contour level of [0.66, 4.62, 8.58, 12.54, 16.5, 20.46, 24.42] $\rm K \  km \ s^{-1} $, and the black dashed contour is the Urumqi $\lambda$  6 cm radio emission with the same contour level of Fig. \ref{Fig:allvelocity}. Lower-right: The latitude-velocity diagram of the whole area (yellow-filled contour) and MC G150.6+03.7 (blue-filled contour). The yellow-filled contour level is [-0.5, 0.2, 1.2, 2.2, 3.2, 4.2, 5.2, 6.2, 7.2, 8.2, 9.2, 12] K, and the blue-filled contour level is [0.2, 1.2, 2.2, 3.2, 4.2, 5.2] K. Upper-left: The longitude-velocity diagram with the same colour and level setting of the latitude-velocity diagram. Upper-right: Average \twco (Blue) and \thco (Red) spectra extracted from the MC G150.6+03.7 region. }
\label{Fig:G150.6+03.7}
\end{figure}

We search for molecular clouds in the [-14, -2] velocity component using the similar method as described in \citet{yan2021distances} with \twco data. This method employs the \textsc{DBSCAN} algorithm (\textit{density-based spatial clustering of applications with noise}; \citealt{ester1996density}), a non-parametric density-based clustering technique. Using the density-based clustering method, molecular clouds can be identified by grouping pixels in the position-position-velocity (PPV) space.

\textsc{DBSCAN} has two key parameters, MinPts and $\epsilon$. MinPts defines the minimum number of pixels required to form a dense region, while the $\epsilon$ parameter sets the radius within which pixels are considered neighbors. Following \citet{yan2021distances}, we adopt the same parameters $\epsilon=1$ and Minpts$=4$, and other four criteria: (1) the minimum voxel number is 16; (2) the minimum peak brightness temperature is 5$\sigma$; (3) the projection area contains a beam (a compact 2$\times$2 region); (4) the minimum channel number in the velocity axis is 3. It should be noticed that criteria (1) and (2) are related to sensitivity, while (3) and (4) are related to resolution. The minimum cut-off on the PPV data cubes is 2$\sigma$ ($\sim 1$ K).

After filtering out regions that cover only a small fraction of the cube area, \textsc{DBSCAN} identified 112 objects. Given the physical size of molecular clouds, we imposed a minimum angular area threshold of $>$100 arcmin$^2$ (equivalent to 400 pixels or $\sim 5 \rm \  pc^{2}$ at a distance of 740 pc) to distinguish between clouds and clumps. As a result, we identified five molecular clouds in the velocity range of [-14, -2] $\rm km \ s^{-1}$(see Table \ref{tab:cloud}).  

We then subtract the masks of each cloud, enabling us to plot their intensity map, l-v diagram, b-v diagram, and mean spectra separately. Figure \ref{Fig:G150.6+03.7} displays the morphology of the largest cloud, MC G150.6+03.7, in both position-position (l-b) space and position-velocity (l-v and b-v) space. A series of figures for the other four clouds are presented in Appendix A.


\subsubsection{Properties of Clouds}
\label{sec:calculation}

\begin{table*}[ht]
\centering
\caption{The Properties of the molecular clouds around the SNR G150.3+4.5}
\label{tab:cloud}
\begin{threeparttable}
\begin{tabular}{C{3.5cm} C{3cm} C{2.5cm} C{3cm} C{2.5cm}}
\hline\hline
Cloud & Column density$^{a}$ & Gas mass$^{a}$ & Column density$^{b}$ & Gas mass$^{b}$\\
Name & ($\times$10$^{20}$ cm$^{-2}$) & ( M$_{Sun}$) & ($\times$10$^{20}$ cm$^{-2}$) & ( M$_{Sun}$)
\\    
\hline
MC G150.6+03.7 & 7.22 & 5229 & 4.75 & 443 \\
MC G151.9+05.2 & 5.58 & 3277 & 5.72 & 514 \\
MC G149.5+04.9 & 3.01 & 182 & 3.19 & 5 \\
MC G151.0+04.6 & 4.57 & 812 & 3.58 & 13 \\
MC G150.9+05.1 & 2.82 & 273 & 3.91 & 19 \\
\hline
\end{tabular}
\begin{tablenotes}
\item[a]{Results estimated from the MWISP $^{12}$CO observations and a CO-to-H2 factor of 2.0$\times$10$^{20}$ cm$^{-2}$ (K km s$^{-1}$)$^{-1}$  \citep{bolatto2013co}. Calculate with distance of 740 pc.}
\item[b]{Results estimated from the MWISP $^{13}$CO observations, assuming LTE condition. Calculate with distance of 740pc.}
\end{tablenotes}
\end{threeparttable}
\end{table*}

With the measured distance at 740 pc and the molecular cloud structure identified by \textsc{DBSCAN}, we then use two methods to calculate the basic physical properties of the molecular clouds based on the \twco and the \thco emission. 

The first method is to use the X-factor ($\rm X_{CO}$) to convert the \twco integrated intensity into $\rm H_{2}$ column density,
\begin{equation}
N_{\rm H_2}=X_{\rm CO} \times \int{T_{\rm MB,^{12}CO}  dV} \ ,
\end{equation}
where $ T_{ \rm MB}$ is the main beam brightness temperature of the corresponding emission lines. The value we adopt for the X-factor is $X_{\rm CO}$ = 
$\rm 2.0\times \ 10^{20} \rm cm^{-2}\  K^{-1}\ km^{-1}\ s$  \citep{bolatto2013co}.

The second method is based on the assumption that the clouds are under the local thermodynamic equilibrium (LTE) condition and the \twco emission is optically thick as described in previous works (e.g. \citealt{dickman1978ratio,garden1991spectroscopic,guo2021wide,sun2024magnetically}). We first derive the excitation temperature by using the peak intensity of the \twco emission: 
\begin{equation}
T_{\rm ex} = T_{\rm 0}[\ln(1+\frac{T_0}{T_{\rm MB,^{12}CO}+0.819})]^{-1}  \ ,
\end{equation}
where $T_0=$$h\nu/k_{\rm B} $ is the intrinsic temperature, $h$ is the Planck constant and $k_{\rm B}$ is the Boltzmann constant. 

We calculate the opacity and the \thco column density with the \thco line parameters and the excitation temperature, following \cite{bourke1997discovery}:
\begin{equation}
 \tau_{13}=-\ln \left[1-\frac{ T_{\rm MB,^{13}CO} }{T_0}\left(\left[e^{T_0 / T_{\mathrm{\rm ex}}}-1\right]^{-1}-0.164\right)^{-1}\right] \ ,
\end{equation}
\begin{equation}
N_{\rm ^{13}CO}=2.42 \times 10^{14} \frac{\int{T_{\rm MB,^{13}CO} dV}}{1-e^\frac{-T_0}{T_{\rm ex}}} \times \frac{\tau_{13}}{1-e^{-\tau_{13}}} \ \ .
\end{equation}

To derive the $\rm H_2$ column density traced by $^{13}$CO, we then calculate $R_\text{12/13}$, the isotopic ratio of $^{12}$CO to $^{13}$CO, using the relation $\rm [^{12}C/^{13}C] = 4.08$$D_{\rm GC} + 18.8$, as described in \cite{sun2024improved}. The Galactocentric distance $D_{\rm GC}$ is 8.95 kpc, based on the longitude at 150\degree and the heliocentric distance at 740 pc, resulting in an estimated $R_\text{12/13} \approx 55$. Given the abundance ratios [H$_2$]/[$^{12}$CO] $\sim 1.1 \times 10^{4}$ \citep{frerking1982relationship}, the abundance ratio [H$_2$]/[$^{13}$CO] is estimated to be $6\times10^{5}$. The column density of molecular hydrogen $N_{\rm H_2}$ is converted from $N_{\rm ^{13}CO}$ with this ratio.

To calculate the cloud mass, we use the mask derived from \textsc{DBSCAN} to generate the moment 0 maps (integrated intensity maps) of both \twco and \thco. We then calculate the total cloud masses by integrating the $\rm H_{2}$ column density over the area of the \twco and \thco emissions within the cloud boundary mask, applying the following formula:
\begin{equation}
\label{equ:mass calculate}
M=\mu m_{\rm H}D^{2}\int{N(\rm H_{2})  d\Omega} \ ,
\end{equation}
where $\mu$ is the mean molecular weight per hydrogen molecule (assumed to be 2.83, \citealt{kauffmann2008mambo}), $\rm m_{H}$ is the mass of atomic hydrogen, D is the distance to the object, and $\rm d\Omega$ is the solid angle element. We first calculate the mass of each pixel with Equation (\ref{equ:mass calculate}), and then sum all the pixels to obtain the total gas mass $M_{\rm {X_{CO}}}$ and $M_{\rm {LTE}}$ of the molecular clouds, respectively. All these results are listed in Table \ref{tab:cloud}. A further detailed description of these clouds are discussed in Appendix A. In summary, we derive a total molecular gas mass of $\sim 10^4\  \rm M_{\odot}$ in the SNR region. 


\subsection{The Association between Molecular clouds and SNR G150.3+4.5} 
\label{sec:association}
\label{sec: morphology association}
\label{sec: kinematic}

\begin{figure}[h!]
\resizebox{\hsize}{!}{\includegraphics{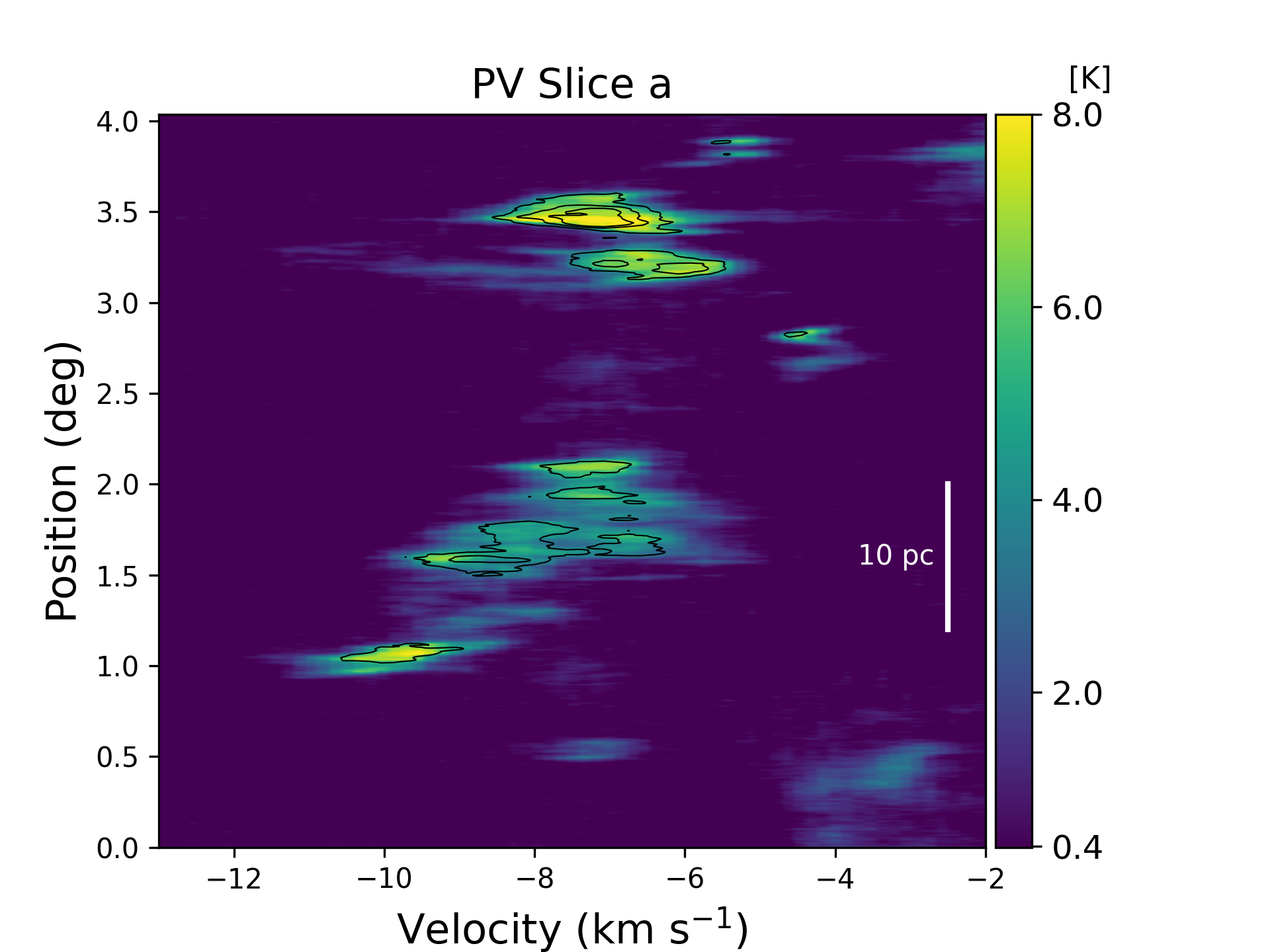}}
\caption{ The position-velocity (pv) map of  \twco along the direction marked in Fig. \ref{Fig:m1}. The contour shows the \thco emission with the contour level of [1.3,2.6,3.9] K ($\rm 1 \sigma = 0.13 K$), while the background shows the \twco emission.} 
\label{Fig:pv}
\end{figure}

\begin{figure*}[h!]
\resizebox{\hsize}{!}{\includegraphics{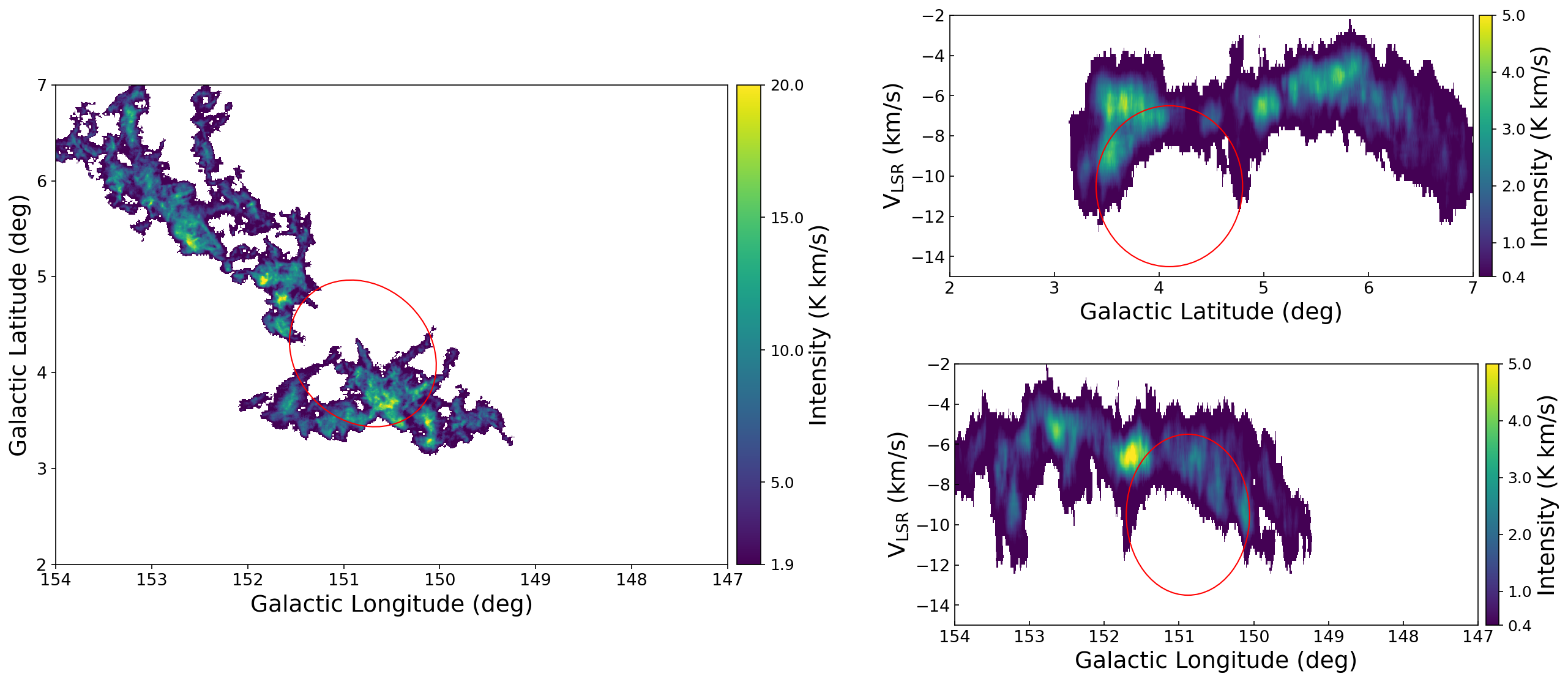}}
\caption{Left panel: Integrated intensity images for Clouds: MC G150.6+03.7, MC G149.5+04.9, MC G151.9+05.2, and MC 150.9+05.1. Right panels: The l-v (top) and b-v (bottom) diagrams show the \twco data integrated over the selected clouds. An approximate fitting is marked by a red solid-line ellipse, illustrating an arc-like structure in the diagrams. }
\label{Fig:expand}
\end{figure*}

The CO emission of the [-14, -2] $\rm km \ s^{-1}$ component is spatially distributed around the radio loop of SNR G150.3+4.5, suggesting the potential association between the gas and SNR. We present the intensity-weighted velocity map of \twco in $\rm Fig.\  \ref{Fig:m1}$. The two distinct $V_{\rm LSR}$ values of -8 $\rm km \ s^{-1}$ and -6 $\rm km \ s^{-1}$ are observed in the southwest and northeast, respectively. 
However, considering that the eastern cloud MC G151.9+5.2 and the southern cloud MC G150.6+3.7 are at the same distance, it is more likely that they are related, rather than being a mere projection coincidence of two unrelated clouds.

In Fig. \ref{Fig:pv}, we present the position-velocity (PV) diagram of \twco and \thco emission along the directions indicated by the black arrows shown in Fig. \ref{Fig:m1}. We found an arc-like structure in the PV diagram. The arc-like structure spans a velocity range from -12 to -5 $\rm km \ s^{-1}$ and exhibits significant velocity dispersion. The prominent broad-line wing of the \twco is from the northeast cloud MC G151.9+5.2, where is the Position C in $\rm Fig.\  \ref{Fig:m1}$. The black contour shows the \thco emission, which also exhibits an arc-like distribution in the PV diagram.

In Fig. \ref{Fig:expand}, we overlap the b-v diagram of the clouds identified in Sec. \ref{sec:clouds}. There is also an arc-like structure across the remnant in the b-v diagram. It can be fitted by an ellipse with a 4 $\rm km \ s^{-1}$ half-axis length. A similar arc-like structure is also found in the l-v diagram. The ellipse center of the b-v and l-v diagrams are \textit{b} $\approx$ 4\fdg1 and \textit{l} $\approx$ 150\fdg9, respectively, which is close to the center of the SNR. 
Therefore, we use an ellipse with its center located at (150\fdg9, 4\fdg1), a major axis of 0.82 degrees in length, and a minor axis of 0.7 degrees in length to fit the arc-like structure observed in the l-v or b-v diagram. Additionally, the arc-like structure in the PV diagram (Fig. \ref{Fig:pv}) can also be fitted by an ellipse with the $4 \ \rm km \ s^{-1}$ half-axis and a similar center position. The arc-like structures suggest that the clouds are expanding with an expanding velocity of $v_{\rm exp} \thicksim 4 \ \rm km \ s^{-1}$.

Such expanding gas motions have also been observed in other SNR-MC systems with similar expansion velocities, e.g., Kes79 ($4 \ \rm km \ s^{-1}$; \citealt{kuriki2018discovery}), RX J0046.5-7308 ($3 \ \rm km \ s^{-1}$; \citealt{sano2019discovery}), and W49B ($6 \ \rm km \ s^{-1}$; \citealt{sano2021alma}). In our case, the presence of the expanding molecular gas shell provides kinematics evidence for the association with the SNR.

However, it is essential to consider alternative explanations for the expansion and the loop-like spatial distribution. One plausible explanation is that the presence of bright stars, characterized by strong stellar winds and intense ultraviolet (UV) radiation, may be responsible for the non-gaussian spectra and for shaping the observed distribution of molecular clouds (see, e.g. \citealt{churchwell2007bubbling}). 
We search for the massive OB stars within this region in the SIMBAD Astronomical Database\footnote{\href{http://simbad.cds.unistra.fr/simbad/}{http://simbad.cds.unistra.fr/simbad/}}.
Considering the estimated distance of 740 pc for the molecular clouds in the [-14, -2] $\rm km \ s^{-1}$ range, we exclude OB stars that are located at either less than 500 pc or more than 1 kpc. The distances come from the Gaia DR3 catalog \citep{bailer2021estimating}.
We find several massive OB stars with distances ranging from 500 pc to 1 kpc, located to the left of the loop and in regions without CO emissions, as shown in Fig. \ref{Fig:aligns}. Notably, a spectral type O9.7 IIn star 1 Cam A (also referred to as HD 28446 A; \citealt{sota2014galactic}) is located close to MC G150.6+03.7 and MC G151.9+05.2. The angular distance between 1 Cam A and the MC G151.9+05.2 is about 1\textdegree . \cite{chen2013linear} found a linear relationship between the radius of a bubble surrounding a main-sequence star in a molecular environment and the stellar mass. For a typical spectral type O9.5 star, the radius of the stellar wind-blown bubble is approximately $\sim$ 11 pc \citep{chen2013linear}, which is smaller than the distance between the star and the prominent disturbing Position C. For star 1 Cam A, the measured distance is 790 pc. Considering that it is situated approximately 30 pc further from the [-14, -2] $\rm km \ s^{-1}$ molecular gas, we suggest that the disturbing gas is not associated with the O9.7 IIn star 1 Cam A.

Based on the morphology alignment and the kinematics evidence, we suggest that the molecular clouds in the velocity range [-14, -2] $\rm km \ s^{-1}$ are associated with the SNR.

\section{Discussion} \label{sec:discussion}

\subsection{Potential Shocked Gas Positions} 
\label{sec:shocked}

\begin{figure*}[h!]
\resizebox{\hsize}{!}{\includegraphics{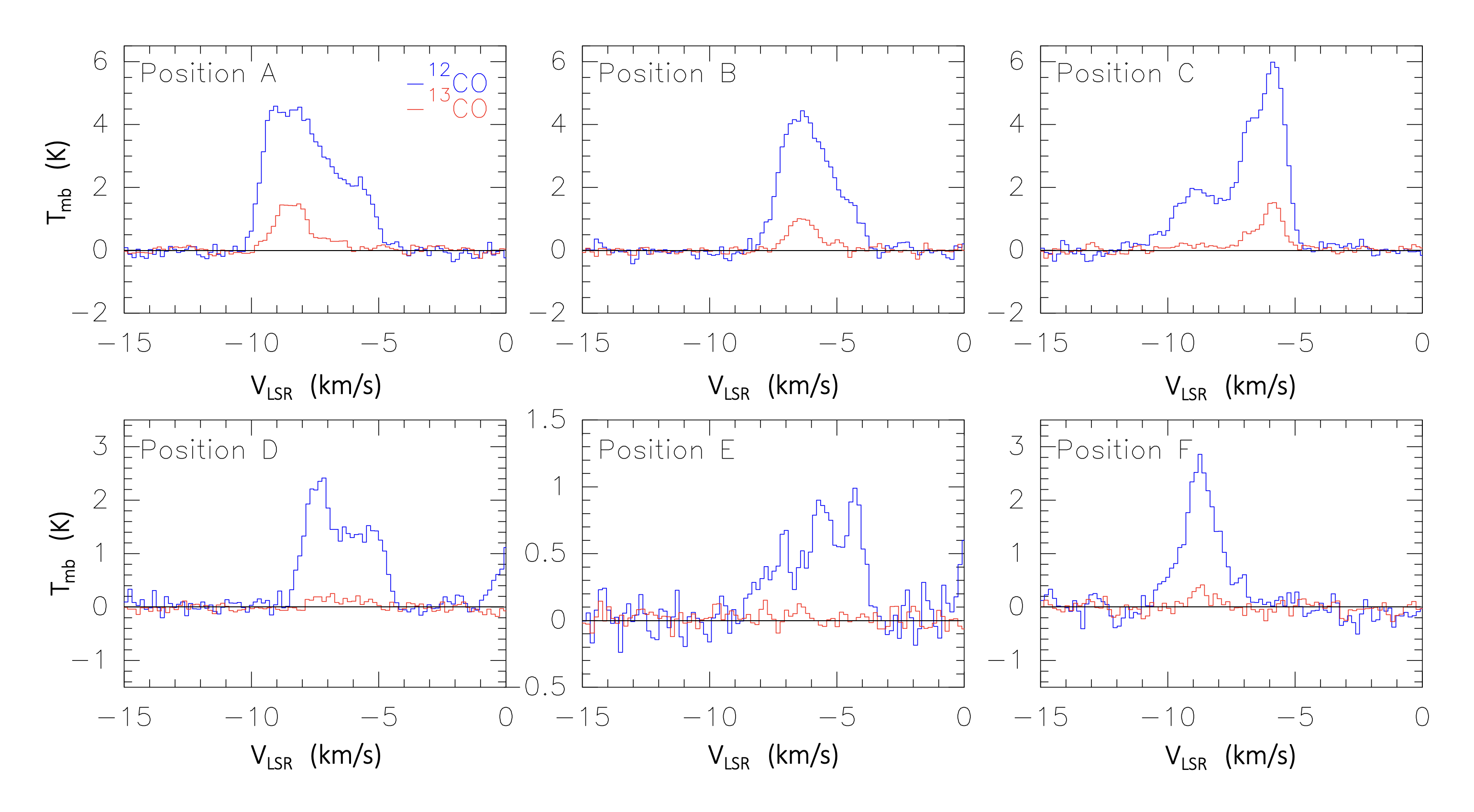}}
\caption{ The CO spectra in the surrounding clouds of the SNR G150.3+4.5. The spectra are sampled from the six position marked in Fig. \ref{Fig:m1}, with an area of 1$'$$\times$1$'$. In each panel, the blue and red spectra represent the emission from the \twco and \thco. The spectra all show large velocity dispersion.}
\label{Fig:spec}
\end{figure*}

\begin{figure*}[h!]
\resizebox{\hsize}{!}{\includegraphics{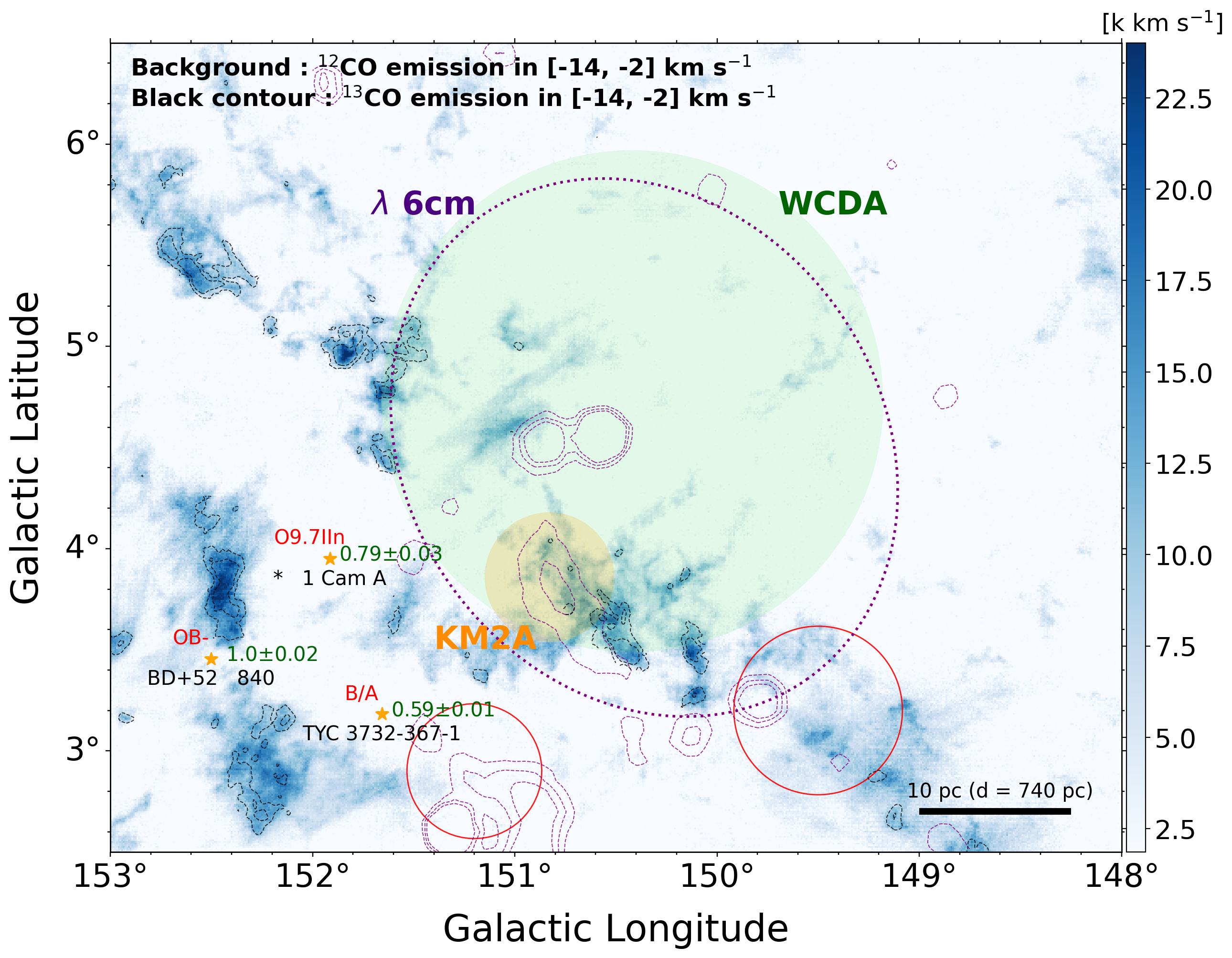}}
\caption{The integrated intensity map of \twco with the color scale range from $3\sigma \ (1.9 \ \rm K \ km \ s^{-1})$. The black contour shows the \thco emission with the contour level of [1.75, 3.5]  ($5\sigma = 1.75$) $\rm K \ km \ s^{-1}$. The velocities are both integrated between [-14, -2] $\rm km \ s^{-1}$. The purple dashed ellipse and contours show the size and intensity with the contour level of [12, 15, 18] K of Urumqi $\lambda$  6 cm radio continuum emission. The orange and green circles represent the LHAASO source 1LHAASO J0428+5531 \citep{cao2023lhaaso}, with orange indicating KM2A components and green denoting WCDA components. The CO emission aligns well with the radio shell and spatially coincident with the LHAASO source. The yellow star markers represent massive OB stars, with the red labels indicating the spectral types and the green labels displaying the distances with a unit of kpc.}
\label{Fig:aligns}
\end{figure*}

\begin{table*}[ht]
\centering
\caption{Spectral Parameters of the Shocked Region}
\label{tab:spec}
\begin{threeparttable}
\begin{tabular}{C{1.5cm} C{3cm} C{3cm} C{3cm} C{2cm} C{2cm}}
\hline\hline
Region & Position & Line & V$_{LSR}$(Peak) & T(Peak) & FWHM$^{b}$ \\
& & & ($\rm km \ s^{-1}$) & (K) & $\rm km \ s^{-1}$ \\
\hline
A & 150.4\degree +3.7\degree & $^{12}$CO (J=1-0) & -8.00 (0.04) & 4.50 & 3.4 \\
& & $^{13}$CO (J=1-0) & -8.40 (0.02) & 1.51 & 1.5 \\
B & 151.5\degree +3.6\degree & $^{12}$CO (J=1-0) & -6.18 (0.02) & 4.29 & 2.4 \\
   &   & $^{13}$CO (J=1-0) & -6.34 (0.03) & 1.02  & 1.3\\
C & 151.6\degree +4.8\degree & $^{12}$CO (J=1-0) & -6.01 (0.02) & 5.33 & 3.3 \\
   &   & $^{13}$CO (J=1-0) & -5.96 (0.03) & 1.34  & 1.2\\
D & 151.4\degree +5.5\degree & $^{12}$CO (J=1-0) & -6.82 (0.05) & 1.70 & 2.5 \\
   &   & $^{13}$CO (J=1-0) & -$^{a}$ & - & - \\
E & 150.5\degree +5.7\degree & $^{12}$CO (J=1-0) & -5.51 (0.14) & 0.64 & 2.8 \\  
   &   & $^{13}$CO (J=1-0) & -   & -  & -\\  
F & 149.5\degree +4.9\degree & $^{12}$CO (J=1-0) & -8.69 (0.14) & 2.48 & 1.7 \\  
   &   & $^{13}$CO (J=1-0) & -   & -  & -\\  
\hline
\end{tabular}
\begin{tablenotes}
\item[a] No $^{13}$CO emission detected.
\item[b] The Full Width at Half Maximum (FWHM) of the spectra.
\end{tablenotes}
\end{threeparttable}
\end{table*}


Shocks interacting with dense molecular gas are particularly interesting since they influence the chemistry and structure of potential star-forming material and are likely locations for cosmic-ray acceleration. These shocks result in broadened molecular line emission. We inspect the spectra of gas emissions within the rectangular region shown in Fig. \ref{fig:global view} to search for the potential shocked regions. 

We find several regions ($1' \times 1'$) with non-Gaussian deviation exhibiting line broadening profiles in spectra, as shown in $\rm Fig.\  \ref{Fig:spec}$. The locations of these regions are shown in Fig. \ref{Fig:m1} with capital letters, and their spectra profiles are listed in Table \ref{tab:spec}. The prominent broadened line profiles are found in Positions A (MC G150.6+3.7) and C (MC G151.9+5.2), and the fitted full width at half maximum (FWHM) is 3.4 $\rm km \ s^{-1}$ and 3.3 $\rm km \ s ^{-1}$, respectively. In the SNR-MCs interaction systems, a broad FWHM over 6 $\rm km \ s^{-1}$ is typically observed in the \twco spectra (e.g., \citealt{2016ApJ...816....1K,zhou2023systematic}). Some SNR-MCs systems even exhibit dozens of $\rm km \ s ^{-1}$ wing profiles in CO observations (e.g.,  IC443, \citealt{zhang2010co}).

On the other hand, the observation studies of some SNR-MCs systems also show relatively narrow linewidths (2 to 6 $\rm km \ s^{-1}$), such as 3C397 \citep{jiang2010cavity}, N132D \citep{sano2020alma}, and W49B \citep{sano2021alma}. These SNRs all exhibit a wing-like profile, suggesting that narrow linewidths with wing profiles could also be an indicator of shock disturbance. Furthermore, \cite{reynoso2000co} reports three cases: G349.7+0.2, CTB37A, and G16.7+0.1 that have been detected with OH 1720 MHz masers, yet no prominent broadened line profiles (< 3 $\rm km \ s^{-1}$) or significant wing profiles are found in their CO spectra. The presence of OH 1720 MHz masers is a strong indication of SN shock interactions with molecular clouds (see, e.g. \citealt{elitzur1976oh,frail1996survey,frail1998oh,green2002supernova}). Thus, the absence of broadened line profiles cannot exclude the shock scenario without additional observation. In this case, we list these positions here as potential shock candidates.

The spectra of Position A and Position C are similar to these cases with narrow linewidths. Moreover, the \twco spectra of Position A and C exhibit broadening with respect to the \thco spectra ($V_{\rm LSR}$ at $ -6 \ \rm km \ s^{-1}$ for Position A; $V_{\rm LSR}$ at $ -8 \ \rm km \ s^{-1}$ for Position C). Such features often occur in SNR-MCs interacting systems. This is because the broadened profiles of \twco emissions typically originate from turbulent molecular gas that is easily and significantly influenced by local shocks. Conversely, the optically thin \thco emission primarily comes from the quiescent and undisturbed molecular clouds \citep{frail1998oh,2016ApJ...816....1K,su2017hess}. Position A is also located at the edge of the strong radio shell, suggesting a spatial association between the edge of molecular gas and the shock front. Therefore, Positions A and C are very likely to be the shocked regions instead of multi-velocity components. The absence of a broad velocity wing (> 10 $\rm km \ s^{-1}$) might be due to beam dilution from small emitting areas or the sensitivity limitation \citep{enokiya2023discovery}.

As for the other positions, despite their spectra exhibiting non-Gaussian profiles, they show an even narrower linewidth (< 3 $\rm km \ s^{-1}$). Unlike Positions A and C, these positions show no significant \thco emission. These spectra could be affected by shock or explained by multi-velocity components. Further observation is needed to inspect the underlying physics of these positions.

Interstellar magnetohydrodynamic (MHD) shocks are expected only across apatial scales of $\sim 10^{-3} \ \text{pc}$, which is difficult to resolve in current observations \citep{gusdorf2008sio}. Nevertheless, when large-scale flows of molecular gas are pushed to collide, the resulting shocks can span parsec/sub-parsec spatial scales \citep{wu2015gmc,cosentino2019interstellar}. The different features between those larger broadening profiles observed in cases like IC 443 and our study may stem from differences in the ISM environment, including aspects like density, temperature, magnetic field, or spatial distribution. In the future, observations with higher resolution and sensitivity, or studies of high-order transitions of CO combined with large velocity gradient (LVG) analysis, could offer insights into how shocks interact with molecular gas at Positions A and C. These observations may also help determine if shocks influence other positions.



\subsection{Evolutionary stage of SNR G150.3+4.5}
\label{sec:nature}

The association between the molecular clouds and the SNR helps us to obtain the distance of the remnant. 
Therefore, at a distance of 740 pc, the dimension of the radio loop of the remnant is $\thicksim$ 40 pc $\times$ 33 pc (3\fdg0 $\times$ 2\fdg5, \citealt{gao2014discovery}), and the height above the Galactic plane is $\thicksim$ 60 pc.

To further study the nature of SNR G150.3+4.5, we use the formula described in \cite{cox1972cooling}:
\begin{equation}
    R_s = 12.9(\frac{t_{age}}{10^4 \rm yrs})^{0.4}(\frac{E_{SN}}{10^{51} \rm{ ergs}})^{0.2} (\frac{n_0}{\rm cm^{-3}})^{-0.2}  \ .
\label{equ:shock}
\end{equation}
\begin{equation}
    v_s = 505(\frac{t_{age}}{10^4 \rm yrs})^{-0.6}(\frac{E_{SN}}{10^{51} \rm{ ergs}})^{0.2} (\frac{n_0}{\rm cm^{-3}})^{-0.2}  \ .
\end{equation}
It should be noted that the assumed explosion energy and the ambient density could differ by a few orders of magnitudes from their actual values.

Here we attempt to bring out an evolutionary scheme for the SNR based on the gas properties. With a total molecular gas mass of $10^4 \ \rm M_{\odot}$ (the sum of all cloud masses), and considering a spherical volume with a radius of 20 pc and a mean molecular weight of $\mu=2.8$, we use the formula for volume density $\rho = 3M/4\pi r^3$ to derive a lower limit of ambient density $n_0 \thicksim 5 \rm \ cm^{-3}$.
We assume a typical SNe energy $E_{SN}= 10^{51} \ \rm{ ergs}$, together with the ambient density $n_0 = 5 \ \rm cm^{-3}$ and the radius of the shock front is $R_s = 20 $ pc. We estimate the age of the SNR to be $t_{age} \approx 3.8 \times 10^4$ years, with a shock velocity of $v_s \approx 162 \ \rm km \ s^{-1}$. A similar age $t_{age} \approx 2.6 \times 10^4$ years can be derived by the statistical diameter-age (D-t) relation $(\frac{D}{\text{pc}}) = (9.52 \pm 0.92)(\frac{t}{\rm kyr})^{0.44\pm0.04}$ \citep{ranasinghe2023statistical}.

Nevertheless, the molecular gas surrounding the SNR is inhomogeneous. As mentioned in Sec. \ref{sec:results}, the CO emissions are primarily distributed in the southeast of the SNR. This suggest shocks might propagate in a higher ambient density than the average value, which lead to an older age. To derive an upper limit, we further use the HI4PI HI data \citep{bekhti2016hi4pi} to estimate the atom gas mass. With the same velocity range (-14 to -2 $\rm km \ s^{-1}$) in the region of the southern shell (MC G150.6+03.7), we obtain an HI mass of $\sim 11000 \ \rm M_{\odot}$ and thus a total gas mass (H$_2$ and HI) of $\sim 16200 \rm \  M_{\odot}$. The corresponding ambient density is $n_0 \thicksim 110 \ \rm cm^{-3}$. In this case, the upper limit of the SNR age is derived as $\thicksim 18 \times 10^4 $ years, and the shock velocity $v_s\thicksim 35$ $\rm km \ s^{-1}$.

In Sec. \ref{sec:association}, we find an expanding gas shell along with the radio continuum shell with an expansion velocity of $\thicksim$ 4 $\rm km \ s^{-1}$. Such an expanding gas shell could formed by the shock from the SNR, or strong stellar winds from the high-mass progenitor of the SNR. In the wind-blown scenario, the SNR evolution is dominated by the Sedov-Taylor phase when it expands in the low-density wind-blown bubble and then ends immediately when the blast wave hits the dense shell \citep{weaver1977interstellar}. If the SNR evolves in a wind-blown bubble, the evolutionary phase of the remnant should be speeded up, which therefore leads to a lower age \citep{dwarkadas2005evolution,dwarkadas2007evolution}. Assuming an ambient density of $n_0 \thicksim 0.1 \ \rm cm^{-3}$ in a low-density wind bubble \citep{weaver1977interstellar}, the timescale for the shock front to reach the dense gas shell is calculated of $\sim$ 8 kyrs with Equ. \ref{equ:shock}. Considering the absence of significant X-ray emission, the SNR might have lost a significant portion of its explosion energy due to radiative cooling. According to \citep{haid2016supernova}, within approximately 2 kyrs, 80\% of the initial thermal energy is emitted as radiation, and this occurs almost regardless of the shell density. Thus, a lower limit on the order of 10$^4$ years for the age of the SNR is reasonable.

Assuming the progenitor of the remnant was a massive single star and the current molecular shells are primarily the result of its massive wind, we can also estimate the mass of the progenitor based on the size of the wind-blown bubble. Adopting the fitted size of 9 pc (0\fdg7) as the radius of the wind-blown bubble, the initial stellar mass is at least 14 M$_{\odot}$, based on the linear relation $R_{\rm c}({\rm pc})\approx 1.22 M_{\rm star}/M_{\odot}- 9.16 \ \rm pc$ described in \cite{chen2013linear}. This suggests that the spectral type of the progenitor is likely earlier than B2, given a constant interclump pressure of $p/k \sim 10^5 \, \text{cm}^{-3} \text{K}$.
In this scenario, the progenitor of the SNR might be a massive B-type star that exploded in a wind-blown bubble that was swept by its stellar wind. After the SNe, the gas in the formed shell is accelerated by the shock of the SNR.

\cite{devin2020high} suggests a lower limit $t_{age} \approx 1 \times 10^3$ years for the SNR, based on the assumption that the ambient density $n_0 = 3 \times 10^{-3} \ \rm cm^{-3}$ and a distance of 0.7 kpc. In this model, the SNR has not yet reached the Sedov phase and has exploded in a very low-density environment. However, our findings of the association between the molecular gas and the SNR suggest that the ambient density is unlikely to be as low as proposed. Considering that historically recorded SNRs of similar age (1 kyr) are situated at distances greater than 0.7 kpc (e.g., SN 1054 at 2 kpc, \citealt{trimble1968motions}; and SN 1006 at 2.2 kpc, \citealt{winkler2003sn}), the lack of historical records for the SNR also implies it is unlikely to be as young as suggested. Thus, we suggest an age of the SNR $t_{age} \approx 3.8 \times 10^4$ years by assuming a uniform ambient density of $n_0 \thicksim 5 \rm \ cm^{-3}$, and the possible range of the age is (1-18) $\times$ 10$^4$ years.



\subsection{The Origin of TeV Source 1LHAASO J0428+5531}

\begin{figure*}[h!]
\resizebox{\hsize}{!}{\includegraphics{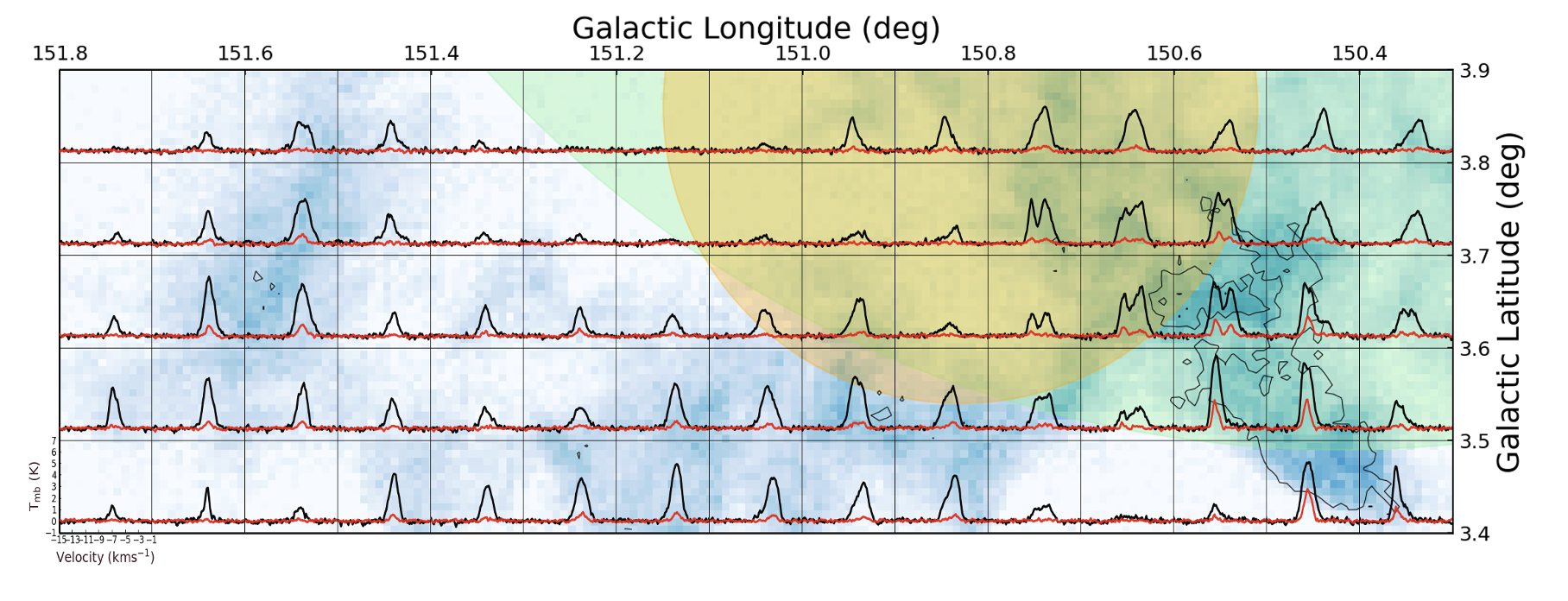}}
\caption{The gridded spectra plot for MC G150.6+03.7. The background shows the intergrated intensity of \twco emission, while the overlaid spectra of both \twco (black) and \thco (red) represent the average spectra for each respective grid area. The green shadow indicates the WCDA component as in Fig. \ref{fig:global view}, while the yellow shadow area indicates the KD2A component.}
\label{Fig:overlaid spec c1}
\end{figure*}

\begin{figure}[h!]
\resizebox{\hsize}{!}{\includegraphics{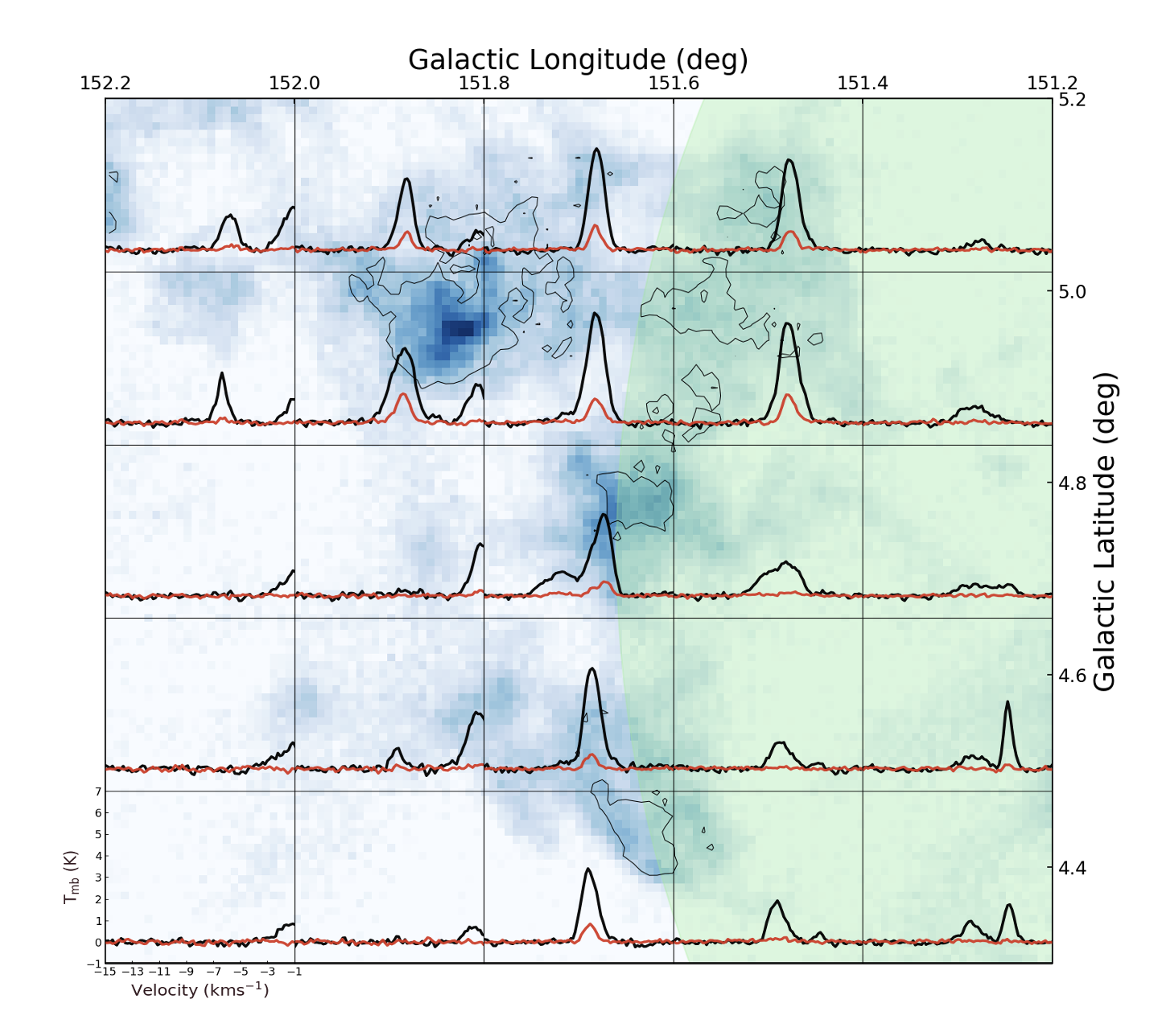}}
\caption{Same as in Fig. \ref{Fig:overlaid spec c1}. The gridded spectra plot for MC G151.9+05.2. }
\label{Fig:overlaid spec c2}
\end{figure}

Previous studies have found several SNR-MCs interacting systems associated with TeV sources, 
such as IC443 \citep{su2014molecular}, W51C \citep{aleksic2012morphological}, W49B \citep{zhou2018asymmetric}, W28 \citep{abdo2010fermi}, and HESS J1912+101 \citep{su2017hess}.  In the case of SNR G150.3+4.5, similar in age to IC443, \cite{zeng2023high} recently discussed two scenarios regarding the origin of TeV Source 1LHAASO J0428+5531. One scenario suggests that the high-energy $\gamma$-ray emission in this region results from the interaction between accelerated particles from the SNR and the surrounding dense medium. 

In Fig. \ref{Fig:aligns}, we present the \thco integrated intensity with the velocity range of [-14, -2] $\rm km \  s^{-1}$. We compare the spatial distribution between \thco emission and the LHAASO Source 1LHAASO J0428+5531 in Fig. \ref{Fig:aligns}. The southern cloud MC G150.6+3.7 in \twco emission decomposes to different structures in $^{13}$CO. The densest part is located around \textit{l} $\approx$ 150\fdg5, \textit{b} $\approx$ 3\fdg5 and shows a shell structure aligned well with the shell in radio emission, while the latter one is also the strongest part of the Urumqi $\lambda$ 6 cm emission with a non-thermal radio spectra. Meanwhile, this CO shell structure coincides with the very high energy (VHE) emission of 1LHAASO J0428+5531 KM2A component. In the grided-spectra plot presented in Fig. \ref{Fig:overlaid spec c1} and \ref{Fig:overlaid spec c2}, we also observed a correlation with the edge of the WCDA component and the likely shocked gas region (the non-Gaussian \twco spectra exhibit broadening with respect to the \thco spectra). Furthermore, we find that the ellipse fitted to suggest expanding molecular gas in Fig. \ref{Fig:expand} coincides with the KM2A component of the LHAASO source 1LHAASO J0428+5531. This overlapping alignment suggests an association between the expanding shocked gas and 1LHAASO J0428+5531.  

Combining these spatial coincidence between KM2A/WCDA components of 1LHAASO J0428+5531 and the dense molecular gas traced by CO, and the evidence of SNR-MCs interaction mentioned in Sect. \ref{sec:association}, we suggest that the VHE emission of 1LHAASO J0428+5531 comes from the hadronic origin of the SNR-MCs interaction, as discussed by \cite{zeng2023high}.

High-resolution observation of the CO high-order transition towards IC443 presented by \cite{dell2020interstellar}, suggests that the shocked clump with the most prominent broadening spectra only holds a fraction ($\sim 200 \ \rm M_{\odot}$) of the total molecular mass ($\sim 10^{3} \ \rm M_{\odot}$) in the extended G region. The surrounding dense structure should be taken into account to investigate the origination of CRs. In our study, Position C might be the main target similar to the shocked clump in the extended G region of IC443 due to its clumpy morphology and the shocked spectral profiles as mentioned in Sec. \ref{sec:shocked}. However, Position C can only attribute to the WCDA component based on the current LHAASO data. Position A could attribute to both KM2A and WCDA components, while it shows a more complicated structure. In general, the radio emission is also related to relativistic particles, while the radio emission traces shocks into gas with lower density, which do not lead to acceleration to TeV energies. Nevertheless, the spatial alignment of strong radio shell and the KM2A component, as shown in Fig. \ref{Fig:aligns}, suggests a scenario where the dense shocked molecular clump, from which the CRs originate, is embedded in a lower-density environment. Still, it should be noted that the angular resolution for the LHAASO detector is limited, which ranges from 0\fdg5 at 20 TeV to 0\fdg2 at 100 TeV, with data collection still in progress. In the future, more information of the VHE sources and higher resolution observation of the molecular lines, could enhance our understanding of the origins of CRs and the environments from which they arise.


\section{Summary} \label{subsec:tables}
We present large-field CO line observations toward SNR G150.3+4.5 using the PMO 13.7 m millimeter telescope. We find that the molecular gas emission in the [-14, -2] $\rm km \ s^{-1}$ range is spatially distributed along the SNR shell detected in the radio continuum observations. In particular, the southern molecular shell is well consistent with the bright radio shell. Further, the line broadening and asymmetries are found in the CO spectra of the clouds. We also find systematic velocity gradients in PV diagrams, which suggests the expansion of clouds with an expanding velocity of $\thicksim 4 \ \rm km \  s^{-1}$. Based on the morphology and kinematic evidence, we suggest that the remnant is associated with the MCs at $[-14, -2]$ $\rm km \  s^{-1}$ range.
Adopting the distance of 740 pc measured to the clouds, we calculate the dimension of the SNR to be $\thicksim$ 40 pc $\times$ 33 pc, with the height above the Galactic plane estimated to be $\thicksim$ 60 pc. Moreover, we suggest the remnant is in the Radiative phase, and the age is estimated to be $(1-18) \times 10^{4} \ \rm years$. The spatial correlation between the high-energy emission of LHAASO source 1LHAASO J0428+5531 and dense molecular gas traced by $^{13}$CO, combined with line broadening of \twco line, suggests that the VHE emission of 1LHAASO J0428+5531 comes from the SNR-MCs interaction.


\begin{acknowledgements}
We would like to express our sincere appreciation to the anonymous referee for the insightful feedback and recommendations, which have significantly enhanced the quality of the manuscript. 

This research made use of the data from the Milky Way Imaging Scroll Painting (MWISP) project, which is a multiline survey in $\rm ^{12}CO$/$\rm ^{13}CO$/$\rm C^{18}O$ along the northern Galactic plane with the PMO 13.7 m telescope. We are grateful to all the members of the MWISP working group, particularly the staff members at the PMO 13.7m telescope, for their long-term support. MWISP was sponsored by the National Key R\&D Program of China with grants 2023YFA1608000, 2017YFA0402701, and the CAS Key Research Program of Frontier Sciences with grant QYZDJ-SSW-SLH047. This work is supported by the National Natural Science Foundation of China (grant No. 12041305). X.C. acknowledges the support  from the Tianchi Talent Program of Xinjiang Uygur Autonomous Region and the support by the CAS International Cooperation Program (grant No. 114332KYSB20190009). This work also made use of data from the European Space Agency (ESA) mission Gaia (https://www.cosmos.esa.int/gaia), processed by the Gaia Data Processing and Analysis Consortium (DPAC, https://www.cosmos.esa.int/web/gaia/dpac/consortium). Funding for the DPAC has been provided by national institutions, in particular the institutions participating in the Gaia Multilateral Agreement.

\end{acknowledgements}


\bibliographystyle{aa} 

\bibliography{myreference}


\begin{appendix} 

\renewcommand{\thesection}{A\arabic{section}}
\section*{Appendix A: Clouds Description}
\label{sec:appendix}
\renewcommand{\thefigure}{A\arabic{figure}}
MC G150.6+03.7 is the biggest cloud in the [-14, -2] velocity component of this region with the mass $\sim$ 5200 $M_{\odot}$ (Fig. \ref{Fig:G150.6+03.7}). The derived properties of the cloud are shown in Table \ref{tab:cloud}. The dense region of this cloud shows a good coincidence with the radio shell. The velocity dispersion tends to become larger from east to west in the longitude-velocity diagram, and from north to south in the latitude-velocity diagram. 
The largest velocity dispersion is about $\thicksim$ 6 $\rm km \ s^{-1}$ at $ l \sim 150^{\circ}$, and $\thicksim$ 7 $\rm km \ s^{-1}$ at $ b \sim 3.5^{\circ}$, respectively. We present a grid spectra map of the densest part of G150.6+03.7 in Fig. \ref{Fig:overlaid spec c1}. The spectra in each cell show the average spectra of the corresponding grid area. The velocity range for both the integrated map and the spectral axis spans from -15 to -1 $\rm km \ s^{-1}$. We observe numerous cells with obvious deviations from a non-Gaussian spectral profile, indicating disturbances.

\begin{figure}[h!]
\resizebox{\hsize}{!}{\includegraphics{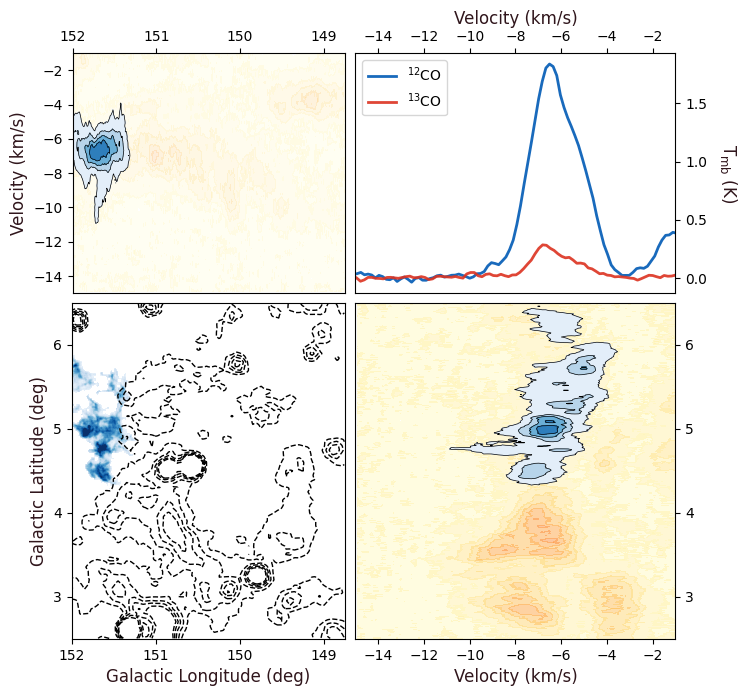}}
\caption{Same as in Fig. \ref{Fig:G150.6+03.7} but for MC G151.9+05.2.  } 
\label{Fig:G151.9+05.2} 
\end{figure}

MC G151.9+05.2 located in the east of the mapping area. There is not much overlap between MC G151.9+05.2 and the radio shell. Both latitude-velocity and longitude-velocity diagrams show a significant spur at position around $l \sim 151.6 , b \sim 4.7$. The CO emission area of MC G151.9+05.2 spatially complements the loop shown in the radio band. As we show in Fig. \ref{Fig:G151.9+05.2}, we drew an ellipse loop based on the outer edge of the radio shell by eyes, and MC G151.9+05.2 has a good coincidence with the loop. However, it should be noticed that this might be just a coincidence because of the subjective eye-based loop. Figure \ref{Fig:overlaid spec c2} presents a grid spectral map of MC G151.9+05.2, similar to Fig. \ref{Fig:overlaid spec c1}. We notice a pronounced line broadening profile in the center of the cloud.   

MC G149.5+04.9 is a filamentary cloud located in the west, and it does not overlap with the radio shell. The emission of MC G149.5+04.9 is weak, as we can notice in the spectra and longitude/latitude-velocity diagram. MC G149.5+04.9 has the lowest surface density of the five clouds.

\begin{figure}[h!]
\resizebox{\hsize}{!}{\includegraphics{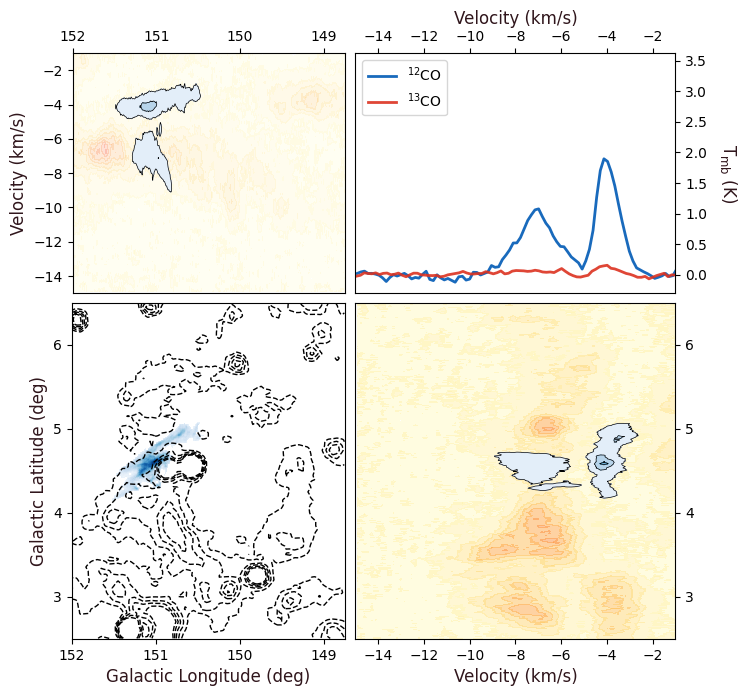}}
\caption{Same as in Fig. \ref{Fig:G150.6+03.7} but for MC G151.0+04.6. }  
\label{Fig:G151.0+04.6}
\end{figure}

MC G151.0+04.6 located next to the radio source in the Urumqi $\lambda$ 6 cm emission (Fig. \ref{Fig:G151.0+04.6}). MC G151.0+04.6 is more like a two-cloud combination rather than a single molecular cloud object. This ambiguity might be due to the \textsc{DBSCAN} parameters setting and the sensitivity and the resolution of our data. The $T_{mb}$ of the boundary between two clouds is higher than the minimum cutoff of \textsc{DBSCAN} ($\sim 1K$), so the algorithm can not separate them. However, the value of the cutoff parameter is already set in a very low level, so we can not lower the parameter otherwise might affect the identification of other clouds. 
In this paper, we do not intend to change the result subjectively from the algorithm. All the properties listed in Table \ref{tab:cloud} are derived from the whole complex object. The two peaks in the mean \twco spectrum are -7 $\rm km \ s^{-1}$ and -4 $\rm km \ s^{-1}$ individually.

\begin{figure}[h!]
\resizebox{\hsize}{!}{\includegraphics{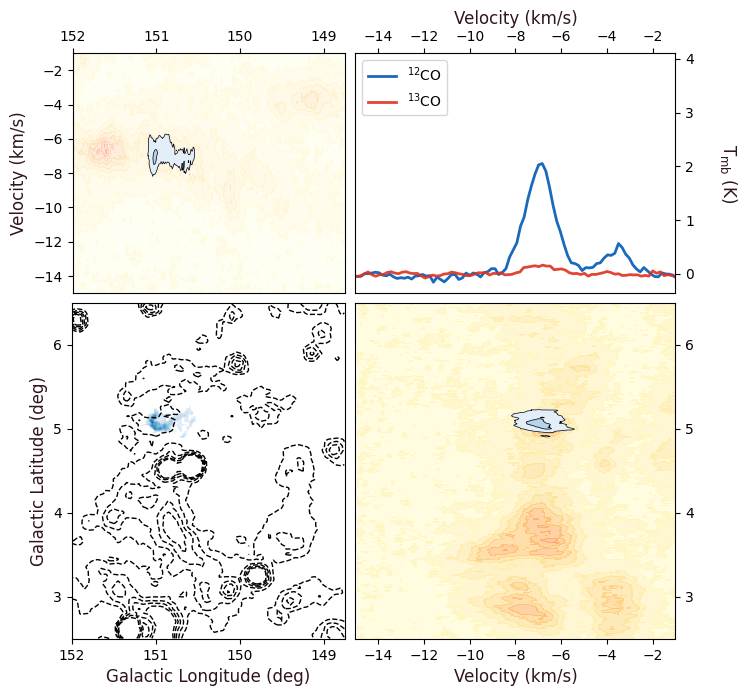}}
\caption{Same as in Fig. \ref{Fig:G150.6+03.7} but for MC G150.9+05.1. }
\label{Fig:G150.9+05.1}
\end{figure}

The \twco emission of MC G150.9+05.1 overlaps with the radio emission between the northern radio shell and the southeastern radio shell (Fig. \ref{Fig:G150.9+05.1}). It shows two peaks in the mean spectrum of \twco emission, at -7 $\rm km \ s^{-1}$ and -3.5 $\rm km \ s^{-1}$ respectively, but only has one peak in the \thco spectrum at -7 $\rm km \ s^{-1}$.


\begin{figure}[h!]
\resizebox{\hsize}{!}{\includegraphics{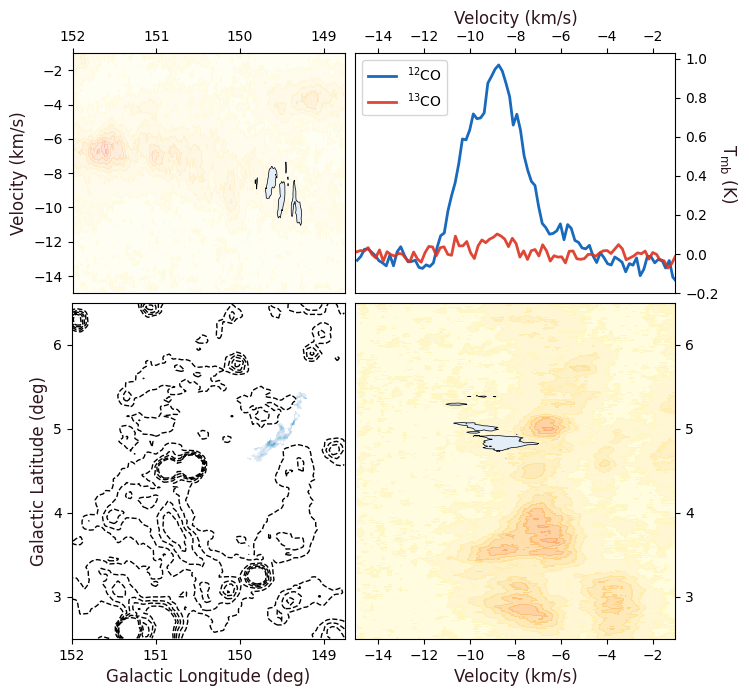}}
\caption{Same as in Fig. \ref{Fig:G150.6+03.7} but for MC G149.5+04.9. }
\label{Fig:G149.5+04.9}
\end{figure}

\end{appendix}

\end{document}